# The MASSIVE Survey - V. Spatially-Resolved Stellar Angular Momentum, Velocity Dispersion, and Higher Moments of the 41 Most Massive Local Early-Type Galaxies

Melanie Veale,[1,2] Chung-Pei Ma,[1] Jens Thomas,[3] Jenny E. Greene,[4] Nicholas J. McConnell,[5] Jonelle Walsh,[6] Jennifer Ito,[1] John P. Blakeslee,[5] Ryan Janish[2]

[1] *Department of Astronomy, University of California, Berkeley, CA 94720, USA*
[2] *Department of Physics, University of California, Berkeley, CA 94720, USA*
[3] *Max Plank-Institute for Extraterrestrial Physics, Giessenbachstr. 1, D-85741 Garching, Germany*
[4] *Department of Astrophysical Sciences, Princeton University, Princeton, NJ 08544, USA*
[5] *Dominion Astrophysical Observatory, NRC Herzberg Institute of Astrophysics, Victoria BC V9E2E7, Canada*
[6] *George P. and Cynthia Woods Mitchell Institute for Fundamental Physics and Astronomy, and Department of Physics and Astronomy, Texas A&M University, College Station, TX 77843, USA*

**ABSTRACT**

We present spatially-resolved two-dimensional stellar kinematics for the 41 most massive early-type galaxies ($M_K \lesssim -25.7$ mag, stellar mass $M^* \gtrsim 10^{11.8}$ $M_\odot$) of the volume-limited ($D < 108$ Mpc) MASSIVE survey. For each galaxy, we obtain high-quality spectra in the wavelength range of 3650 to 5850 Å from the 246-fiber Mitchell integral-field spectrograph (IFS) at McDonald Observatory, covering a $107'' \times 107''$ field of view (often reaching 2 to 3 effective radii). We measure the 2-D spatial distribution of each galaxy's angular momentum ($\lambda$ and fast or slow rotator status), velocity dispersion ($\sigma$), and higher-order non-Gaussian velocity features (Gauss-Hermite moments $h_3$ to $h_6$). Our sample contains a high fraction ($\sim 80\%$) of slow and non-rotators with $\lambda \lesssim 0.2$. When combined with the lower-mass ETGs in the ATLAS$^{3D}$ survey, we find the fraction of slow-rotators to increase dramatically with galaxy mass, reaching $\sim 50\%$ at $M_K \sim -25.5$ mag and $\sim 90\%$ at $M_K \lesssim -26$ mag. All of our fast rotators show a clear anti-correlation between $h_3$ and $V/\sigma$, and the slope of the anti-correlation is steeper in more round galaxies. The radial profiles of $\sigma$ show a clear luminosity and environmental dependence: the 12 most luminous galaxies in our sample ($M_K \lesssim -26$ mag) are all brightest cluster/group galaxies (except NGC 4874) and all have rising or nearly flat $\sigma$ profiles, whereas five of the seven "isolated" galaxies are all fainter than $M_K = -25.8$ mag and have falling $\sigma$. All of our galaxies have positive average $h_4$; the most luminous galaxies have average $h_4 \sim 0.05$ while less luminous galaxies have a range of values between 0 and 0.05. Most of our galaxies show positive radial gradients in $h_4$, and those galaxies also tend to have rising $\sigma$ profiles. We discuss the implications for the relationship among dynamical mass, $\sigma$, $h_4$, and velocity anisotropy for these massive galaxies.

## 1 INTRODUCTION

The most massive galaxies in the local universe represent some of the most evolved galaxies, having the oldest stellar populations and thus the longest potential history for major and minor merger events. This makes them an excellent probe of galaxy evolution at all stages. Stellar kinematic information is one key ingredient for understanding their structure and evolution.

Early long-slit spectroscopic observations of early-type galaxies (ETGs) revealed the kinematic diversity of such galaxies despite their homogeneous photometric appearance (e.g. Davies et al. 1983; Franx & Illingworth 1990; Bender et al. 1994; Fisher 1997). Kormendy & Bender (1996) classified elliptical galaxies as either disky or boxy, rather than by flattening alone, since the observed flattening is mostly driven by inclination. Disky elliptical galaxies are generally fast-rotating and have power-law central light profiles, whereas boxy elliptical galaxies are slow-rotating and have shallow cored central light profiles. Some analyses of these observations also went beyond measuring the velocity $V$ and dispersion $\sigma$ and quantified the asymmetric and symmetric non-Gaussian features in the line-of-sight velocity distribution (LOSVD) using Gauss-Hermite parameters $h_3$ and $h_4$ (van der Marel & Franx 1993; Bender et al. 1994; Fisher 1997). Deviations of up to 10% from a Gaussian LOSVD were found to be common and to be related to the kinematic structure of the galaxies. For example, $h_3$ is anti-correlated with line-of-sight velocity $V$ due to projection effects in many fast rotating galaxies (Bender et al. 1994; Chung & Bureau 2004; Bureau & Athanassoula 2005).

More recently, integral field spectrographs (IFSs) have significantly expanded the earlier 1D long-slit observations by providing detailed 2D maps of stellar and gas velocities within galaxies (e.g., see review by Cappellari 2016). Results from IFS surveys of local galaxies such as SAURON (Emsellem et al. 2004), ATLAS$^{3D}$ (Cappellari et al. 2011a), VENGA/VIXENS (Blanc et al. 2013), SAMI (Croom et al. 2012), CALIFA (Sánchez et al. 2012), and MaNGA (Bundy et al. 2015) support the divide between (boxy) slow rotators and (disky) fast rotators for ETGs, where slow rotators tend to be more massive, more round, and more likely to host kinematically misaligned or distinct components.

The classification of slow and fast rotators has been connected to galaxy merging histories and cosmological structure formation in many studies (e.g., Bendo & Barnes 2000; Jesseit et al. 2007; Bois et al. 2011; Khochfar et al. 2011; Forbes et al. 2016). The most massive slow-rotating galaxies appear to be an end point of galaxy evolution, for galaxies that have ceased in-situ star formation and undergone at least one major dry merger, while fast rotators represent an earlier stage of evolution. Simulations show that on average $\sim 80\%$ of stars in massive galaxies with $M^* \approx 10^{12}\ M_\odot$ are born ex-situ and then accreted onto the galaxies via mergers, while $\sim 90\%$ of stars in Milky Way-sized galaxies are born via in-situ star formation (e.g., Rodriguez-Gomez et al. 2016). Within bins of stellar mass, slow-rotating galaxies have a higher fraction of ex-situ stars than fast rotating galaxies. Several surveys mentioned above (e.g. CALIFA, MaNGA, SAMI) will be able to provide observational constraints on in-situ vs ex-situ star formation.

Despite the numerous surveys of ETGs, massive ETGs with $M^* \gtrsim 10^{11.5}\ M_\odot$ have not been well studied. ATLAS$^{3D}$ , for instance, is volume limited to a distance of 42 Mpc, and only 6 of the 260 galaxies in their sample have $M^*$ above this value, and only 36 are slow rotators (Emsellem et al. 2011). The SLUGGS survey (Brodie et al. 2014) observes a subsample of 25 galaxies from ATLAS$^{3D}$ to much larger radii (up to $\sim 4R_e$) and finds that the kinematic properties of stars near the center do not necessarily correspond with those in the outskirts of the galaxies (Arnold et al. 2014; Foster et al. 2016). Several kinematic studies of ETGs have targeted brightest cluster galaxies (BCGs). Loubser et al. (2008) present radial profiles for $V$ and $\sigma$ from long-slit observations of 41 BCGs, most of which they classify as dispersion supported, and which also show a variety of dispersion profile shapes. IFS studies of BCGs find a high fraction of slow rotators: 3/4 BCGs in Brough et al. (2011) and 7/10 BCGs in Jimmy et al. (2013). Companion galaxies of BCGs in these studies tend to be fast rotators of lesser mass. A better-defined and larger galaxy sample would be needed to assess the extent to which the kinematic differences in these galaxies are driven by galaxy mass, environment (e.g., centrals vs satellites, halo mass, large-scale density), or other factors.

We designed the volume-limited and $M^*$-selected MASSIVE survey to systematically investigate the high-mass regime that was little explored in previous surveys (Ma et al. 2014, Paper I of the MASSIVE survey). These galaxies are likely to host the most massive black holes, most extreme stellar initial mass funcitons, and most dramatic size evolution over cosmic time. The survey targets the 116 most massive galaxies in the northern sky within a distance of 108 Mpc. The survey is complete to an absolute K-band magnitude $M_K < -25.3$ mag, corresponding to a stellar mass of $M^* \gtrsim 10^{11.5}\ M_\odot$. The MASSIVE galaxies are observed with a 107″ square IFS that extends to a few $R_e$ for most galaxies. We reported our first results on the spatial gradients of stellar populations of MASSIVE galaxies using stacked spectra in Paper II (Greene et al. 2015). Paper III (Davis et al. 2016) presented the detections and properties of CO molecular gas in 10 of 15 MASSIVE galaxies from our pilot study. Paper IV (Goulding et al. 2016) analyzed the hot X-ray gas properties of 33 MASSIVE and 41 ATLAS$^{3D}$ galaxies that have archival Chandra X-ray observations.

This paper, Paper V of the MASSIVE survey, presents the first set of results on stellar kinematics for the 41 most massive galaxies, or all galaxies with $M_K \lesssim -25.7$ mag ($M^* \gtrsim 10^{11.8}\ M_\odot$) in the survey. We compare the angular momentum properties and behavior of the $h_3$ parameter to results from the ATLAS$^{3D}$ and SLUGGS surveys, including investigating how angular momentum relates to mass, morphology, and environment. We also study in detail the velocity dispersion profiles and behavior of the $h_4$ parameter, taking advantage of the large radial extent of our data to characterize a variety of profiles that both rise and fall at large radius. Including analysis of the $h_4$ parameter allows us to examine the connections among $\sigma$, $h_4$, dark matter halo mass, velocity anisotropy, and environment.

In Section 2 we describe the data set, and observations. In Section 3 we explain the fitting procedures for the kinematic analysis and present a summary of results. In Section 4 we present more detailed results for velocity and angular momentum, followed by velocity dispersion in Section 5 and higher moments in Section 6. We discuss implications of our results to mass modeling in Section 7, and Section 8 summarizes and concludes.

We also include four appendices. Measurements of central velocity dispersion are compared to literature values in Appendix A. Comparisons of a few individual galaxies to existing literature data are presented in Appendix B. Some technical details of classifying the velocity dispersion profiles are contained in Appendix C. Tables containing detailed properties and results for all 41 galaxies are contained in Appendix D.

## 2 GALAXY SAMPLE AND DATA

This paper presents the stellar kinematics from the Mitchell IFS at the McDonald Observatory for the 41 most luminous ETGs in the MASSIVE survey. These galaxies have $M_K \lesssim -25.7$ mag, which corresponds to stellar masses $M^* \gtrsim 10^{11.8}\ M_\odot$. The full MASSIVE survey is designed to be volume-limited ($D < 108$ Mpc) and complete down to $M^* \approx 10^{11.5}\ M_\odot$ (i.e. $M_K < -25.3$ mag). The survey consists of 116 ETGs selected from the Extended Source Catalog (XSC; Jarrett et al. 2000) of the Two Micron All Sky Survey (2MASS; Skrutskie et al. 2006). The following is a brief summary of the observations and resulting data set. More details can be found in Ma et al. (2014).

### 2.1 Galaxy Properties

Table 1 summarizes the properties of the subsample of 41 MASSIVE galaxies studied in this paper from Ma et al. (2014). The distances are obtained by the surface-brightness fluctuation (SBF) method (e.g. Blakeslee et al. 2009, 2010) for the four galaxies in the Virgo (Blakeslee et al. 2009) or Coma Cluster (Blakeslee 2013). For the others, we use group distances from the High Density Contrast (HDC) catalog (Crook et al. 2007) based on the 2MASS Galaxy Redshift Survey (2MRS) (Huchra et al. 2012). If neither SBF nor HDC distances are available, we apply the same flow model used by the HDC (Mould et al. 2000). We determine the total absolute $K$-band magnitude $M_K$ from equation 1 of Ma et al. (2014), which uses the 2MASS XSC $K$-band magnitude, galactic extinction $A_V$ of Schlafly & Finkbeiner (2011), and the distance described above. Stellar mass is computed from $M_K$ using equation 2 from Ma et al. (2014), which was based on Cappellari (2013):

$$\log_{10}(M^*) = 10.58 - 0.44\,(M_K + 23)\ . \qquad (1)$$

Photometric data are available for 32 of the 41 galaxies in the NASA-Sloan Atlas (NSA, http://www.nsatlas.org) based on the SDSS DR8 catalog (York et al. 2000; Aihara et al. 2011). The effective radius $R_e$, ellipticity $\varepsilon$, and position angle (PA) listed in Table 1 are taken from NSA for these 32 galaxies. For the remaining galaxies, we use values from 2MASS XSC, but apply a correction to $R_e$ based on the overall offset between NSA and 2MASS values (Ma et al. 2014, equation 4).

We make further adjustments to the catalog values for three galaxies in our sample: NGC 4472, NGC 1129, and NGC 4874. For NGC 4472, we adopt $\varepsilon = 0.17$ (Emsellem et al. 2011) and $R_e = 177''$ from deep optical observations (Kormendy et al. 2009, circularized using the listed major-axis $R_e = 194''$) in lieu of the NSA values $\varepsilon = 0.09$ and $R_e = 53.9''$. Our values for $R_e$ in the rest of the sample may still be underestimated due to the relative shallowness of both surveys (e.g. Scott et al. 2013; Cappellari et al. 2011a). We will discuss in subsequent sections where this may impact our results, but even a factor of two increase in $R_e$ for all galaxies would not make a significant difference in any conclusions. For NGC 1129, we find the NSA ellipticity ($\varepsilon = 0.04$) to be significantly smaller than $\varepsilon = 0.15$ from our own imaging data using the Canada France Hawaii Telescope (CFHT); we adopt our own value here. Finally, our ongoing analysis of the kinematic axes of MASSIVE galaxies indicates that NGC 1129 and NGC 4874 have misaligned photometric and kinematic axes (see Section 4.4); we use the kinematic axes (0° and 145°, respectively) for bin folding and other analysis.

NGC 7681 (UGC 12620) was listed in the MASSIVE sample in Ma et al. (2014) with $M_K = -25.72$ mag from 2MASS XSC, which would qualify it to be the 42nd galaxy in the sample studied here. A closer inspection, however, shows that this system consists of a close pair of bulges of roughly equal luminosity separated by 3.6″. A third galaxy is 23″ to the northeast, consistent with the the UGC catalog “pair” classification and the cataloged separation of 0.4′. We include the kinematic maps and properties of NGC 7681 in Figure D11 for completeness but otherwise exclude it from this paper.

### 2.2 Observations

Details of the observations of MASSIVE survey galaxies are described in Ma et al. (2014); here we summarize the pertinent information. We observe the survey galaxies using the Mitchell spectrograph (Hill et al. 2008) on the 2.7 m Harlan J. Smith Telescope at McDonald Observatory. The Mitchell Spectrograph is an optical IFS with 4.1″ diameter fibers and a large 107″ square field of view that consists of 246 evenly-spaced fibers with a one-third filling factor.

Each galaxy is observed with three dither positions of equal exposure time to obtain contiguous coverage of the field of view. (Some galaxies have slightly different configurations, which can be seen in the maps in Appendix D.) We interleave a ten-minute exposure on sky and two twenty-minute exposures on target. With this strategy, the science frames for each galaxy constitute approximately 2 hours of total on-source exposure time, and the central fiber typically reaches a signal-to-noise ratio (S/N) above 50. Outer fibers are binned spatially to improve the S/N in the fainter parts of the galaxies, as described in Section 3.1.

The wavelength coverage of our observations spans 3650 to 5850 Å, which includes the Ca H+K region, the G-band region, H$\beta$, the Mg$b$ region, and several Fe absorption features. The instrumental spectral resolution is determined from the arc lamp spectra, consisting of known mercury and cadmium lines. We use the most prominent 8 lines, spaced roughly equally in the wavelength range from 4000 Å to 5800 Å, and fit a Gaussian to each of these lines. The best-fit FWHM is recorded for each line and for each fiber, with typical values of 4.5 Å and variations with wavelength and fiber position of approximately 0.5 Å, corresponding to a velocity dispersion of $\sigma \sim 100$ km s$^{-1}$ in the redder wavelengths of the spectrum and $\sigma \sim 150$ km s$^{-1}$ in the bluer wavelengths. When fitting binned spectra as described in the next section, these instrument resolution values (averaged over each fiber in a bin) are used to match the resolution of the template spectra to that of the data.

**Table 1.** Galaxy sample: properties of the 41 most massive galaxies in the MASSIVE survey

| Galaxy<br>(1) | R.A.<br>[deg]<br>(2) | Dec.<br>[deg]<br>(3) | $D$<br>[Mpc]<br>(4) | $M_K$<br>[mag]<br>(5) | $R_e$<br>[arcsec]<br>(6) | $\varepsilon$<br>(7) | $PA$<br>[deg]<br>(8) | $R_{\rm max}$<br>[$R_e$]<br>(9) | $\lambda_e$<br>(10) | $\sigma_c$<br>[km/s]<br>(11) | $\langle\sigma\rangle_e$<br>[km/s]<br>(12) | Env.<br>(13) |
|---|---|---|---|---|---|---|---|---|---|---|---|---|
| NGC0057 | 3.8787 | 17.3284 | 76.3 | -25.75 | 27.0 | .17 | 41.1 | 2.9 | 0.02 | 289 | 251 | 1 |
| NGC0315 | 14.4538 | 30.3524 | 70.3 | -26.30 | 25.1 | .28 | 42.4 | 2.5 | 0.06 | 348 | 341 | 6**B** |
| NGC0383 | 16.8540 | 32.4126 | 71.3 | -25.81 | 20.5 | .14 | 141.2 | 3.8 | 0.25**F** | 290 | 257$^\dagger$ | 29 |
| NGC0410 | 17.7453 | 33.1520 | 71.3 | -25.90 | 31.6 | .25 | 34.9 | 2.5 | 0.03 | 291 | 247 | 29**B** |
| NGC0507 | 20.9164 | 33.2561 | 69.8 | -25.93 | 38.4 | .09 | 21.9 | 1.5 | 0.05 | 274 | 257 | 35**B** |
| NGC0533 | 21.3808 | 1.7590 | 77.9 | -26.05 | 40.7 | .26 | 51.2 | 1.9 | 0.03 | 280 | 258 | 3**B** |
| NGC0545 | 21.4963 | -1.3402 | 74.0 | -25.83① | 57.8 | .28 | 59.7 | 1.0 | 0.13④ | 249 | 231 | 32**B** |
| NGC0547 | 21.5024 | -1.3451 | 74.0 | -25.83 | 19.7 | .14 | 94.1 | 2.6 | 0.06 | 259 | 232 | 32 |
| NGC0741 | 29.0874 | 5.6289 | 73.9 | -26.06 | 26.9 | .17 | 86.7 | 0.9 | 0.04 | 292 | 289 | 5**B** |
| NGC0777 | 30.0622 | 31.4294 | 72.2 | -25.94 | 18.6 | .17 | 148.4 | 2.3 | 0.05 | 324 | 291 | 7**B** |
| NGC1016 | 39.5815 | 2.1193 | 95.2 | -26.33 | 26.8 | .06 | 40.5 | 2.9 | 0.03 | 286 | 279 | 8**B** |
| NGC1060 | 40.8127 | 32.4250 | 67.4 | -26.00 | 36.9 | .24 | 74.0 | 1.3 | 0.02 | 310 | 271 | 12**B** |
| NGC1132 | 43.2159 | -1.2747 | 97.6 | -25.70 | 30.9 | .37 | 141.3 | 2.5 | 0.06 | 239 | 218 | 3**B** |
| NGC1129 | 43.6141 | 41.5796 | 73.9 | -26.14 | 30.2 | .15② | 46.2③ | 2.5 | 0.12 | 241 | 259 | 33**B** |
| NGC1272 | 49.8387 | 41.4906 | 77.5 | -25.80 | 31.5 | .07 | 160.3 | 2.4 | 0.02 | 285 | 250 | 117 |
| NGC1600 | 67.9161 | -5.0861 | 63.8 | -25.99 | 41.2* | .26* | 10.0* | 1.9 | 0.03 | 346 | 293 | 16**B** |
| NGC2256 | 101.8082 | 74.2365 | 79.4 | -25.87 | 43.8* | .20* | 75.0* | 1.0 | 0.02 | 240 | 259 | 10**B** |
| NGC2274 | 101.8224 | 33.5672 | 73.8 | -25.69 | 28.4* | .10* | 145.0* | 2.6 | 0.07 | 288 | 259 | 6**B** |
| NGC2320 | 106.4251 | 50.5811 | 89.4 | -25.93 | 19.3* | .30* | 140.0* | 1.0 | 0.24**F** | 340 | 298$^\dagger$ | 18**B** |
| NGC2340 | 107.7950 | 50.1747 | 89.4 | -25.90 | 41.9* | .44* | 80.0* | 2.4 | 0.03 | 232 | 235 | 18 |
| NGC2693 | 134.2469 | 51.3474 | 74.4 | -25.76 | 15.4 | .25 | 166.5 | 4.2 | 0.30**F** | 327 | 296$^\dagger$ | 1 |
| NGC2783 | 138.4145 | 29.9929 | 101.4 | -25.72 | 38.2 | .39 | 165.2 | 2.0 | 0.04 | 252 | 264 | 3**B** |
| NGC2832 | 139.9453 | 33.7498 | 105.2 | -26.42 | 21.2 | .31 | 156.2 | 3.0 | 0.07 | 327 | 291 | 4**B** |
| NGC2892 | 143.2205 | 67.6174 | 101.1 | -25.70 | 23.3 | .06 | 138.4 | 3.3 | 0.05 | 237 | 234 | 1 |
| NGC3158 | 153.4605 | 38.7649 | 103.4 | -26.28 | 16.1 | .18 | 152.6 | 4.6 | 0.26**F** | 301 | 289$^\dagger$ | 6**B** |
| NGC3805 | 175.1736 | 20.3430 | 99.4 | -25.69 | 16.5 | .36 | 64.6 | 5.1 | 0.50**F** | 266 | 225$^\dagger$ | 42 |
| NGC3842 | 176.0090 | 19.9498 | 99.4 | -25.91 | 24.2 | .22 | 1.6 | 1.2 | 0.04 | 262 | 231 | 42**B** |
| NGC4073 | 181.1128 | 1.8960 | 91.5 | -26.33 | 23.0 | .32 | 101.3 | 3.3 | 0.02 | 316 | 292 | 10**B** |
| NGC4472 | 187.4450 | 8.0004 | 16.7 | -25.72 | 177.0② | .17② | 155.0* | 1.0 | 0.20**U** | 292 | 258$^\dagger$ | 205**B** |
| NGC4555 | 188.9216 | 26.5230 | 103.6 | -25.92 | 29.8 | .20 | 117.7 | 2.3 | 0.12 | 328 | 277 | 1 |
| NGC4839 | 194.3515 | 27.4977 | 102.0 | -25.85 | 29.2 | .35 | 65.0 | 0.9 | 0.05 | 261 | 275 | 49 |
| NGC4874 | 194.8988 | 27.9594 | 102.0 | -26.18 | 32.0 | .09 | 40.6③ | 2.4 | 0.07 | 251 | 258 | 49 |
| NGC4889 | 195.0338 | 27.9770 | 102.0 | -26.64 | 33.0 | .36 | 80.3 | 2.4 | 0.03 | 370 | 337 | 49**B** |
| NGC4914 | 195.1789 | 37.3153 | 74.5 | -25.72 | 31.3 | .39 | 155.1 | 2.1 | 0.05 | 233 | 225 | 1 |
| NGC5129 | 201.0417 | 13.9765 | 107.5 | -25.92 | 21.8 | .37 | 5.6 | 3.4 | 0.40**F** | 260 | 222$^\dagger$ | 1 |
| UGC10918 | 264.3892 | 11.1217 | 100.2 | -25.75 | 25.2* | .14* | 5.0* | 2.8 | 0.03 | 247 | 249 | 1 |
| NGC7242 | 333.9146 | 37.2987 | 84.4 | -26.34 | 63.3* | .28* | 40.0* | 1.2 | 0.04 | 255 | 283 | 15**B** |
| NGC7265 | 335.6145 | 36.2098 | 82.8 | -25.93 | 31.7* | .22* | 165.0* | 2.5 | 0.04 | 230 | 205 | 21**B** |
| NGC7426 | 344.0119 | 36.3614 | 80.0 | -25.74 | 20.1* | .34* | 70.0* | 3.2 | 0.56**F** | 284 | 219$^\dagger$ | 4**B** |
| NGC7436 | 344.4897 | 26.1500 | 106.6 | -26.16 | 25.0 | .12 | 13.1 | 2.2 | 0.09 | 280 | 263 | 8**B** |
| NGC7556 | 348.9353 | -2.3815 | 103.0 | -25.83 | 26.4 | .25 | 113.8 | 3.0 | 0.05 | 253 | 243 | 4**B** |

Column notes: see Section 2.1 for more details and citations.
**(1)** Galaxy name (in order of increasing right ascension).
**(2)** Right ascension in degrees (J2000.0).
**(3)** Declination in degrees (J2000.0).
**(4)** Distance in Mpc; from SBF method, group distances from the HDC catalog, or using the same flow model as the HDC catalog.
**(5)** Extinction-corrected total absolute K-band magnitude. Use Equation 1 to convert to stellar mass.
**(6)** Effective radius in arcsec from NSA (where available) or 2MASS (indicated by *, and corrected using eq. 4 of Ma et al. 2014).
**(7)** Ellipticity from NSA (where available) or 2MASS (indicated by *).
**(8)** Photometric position angle in degrees East of North from NSA (where available) or 2MASS (indicated by *).
**(9)** Maximum radial extent of our binned data in units of effective radius (see Section 3.1).
**(10)** Angular momentum within $R_e$ from this paper; **F** = fast rotators, **U** = unclassified, others are slow or non-rotators (see Section 4).
**(11)** Velocity dispersion of the central fiber from this paper (see Section 5). Statistical errors (see Section 3.2) range from 4 to 10 km/s.
**(12)** Velocity dispersion within $R_e$ from this paper (see Section 5). Note that this is an average $\sigma$ over bins within $R_e$, not $\sigma$ for a single spectrum of aperture $R_e$; the difference is significant for galaxies with some rotation, indicated by $^\dagger$ (see Figure 7).
**(13)** Number of group members in the 2MRS HDC catalog, with **B** indicating brightest group/cluster galaxy.
① NGC 545 is a close companion of NGC 547. It is not listed in 2MASS but is designated the BCG of Abell 194 with $M_V = -22.98$ mag in Lauer et al. (2007). The two galaxies have similar magnitudes, so for simplicity we use the $M_K$ of NGC 547 for both galaxies.
② NGC 1129 $\varepsilon$ is from our CFHT data; NGC 4472 $R_e$ is from Kormendy et al. (2009) and $\varepsilon$ is from Emsellem et al. (2011) (see text).
③ NGC 1129 and NGC 4874 have substantial kinematic misalignments (see Section 4.4), so we use the kinematic axis (0° for NGC 1129 and 145° for NGC 4874) instead of the photometric PA for folding and other analysis.
④ NGC 545 $\lambda_e$ is likely overestimated due to systemic velocities (see Section 4.4).

### 2.3 Data Reduction

A detailed explanation of the data reduction process can be found in the appendix of Murphy et al. (2011). The Mitchell Spectrograph was formerly called VIRUS-P (Hill et al. 2008), and is referred to as such in Murphy et al. (2011). Here we provide a brief summary.

We use the in-house data reduction pipeline *Vaccine* developed for Mitchell Spectrograph data (Adams et al. 2011). All bias frames from an observation run are first combined into a master bias for that run. All frames (science, sky, and calibration) are then overscan and bias subtracted. The arcs and twilight flats from either dusk or dawn of each night are combined into a master arc and flat for that night. The fiber trace is constructed by fitting a fourth-order polynomial to the peaks of each fiber, and the spectra of each science and sky frame is extracted from a 5-pixel wide aperture around the trace of each fiber. Wavelength solutions are determined for each fiber and for each night by fitting a fourth-order polynomial to known mercury and cadmium arc lamp lines.

Sky frames for each science frame are made from combining the sky frame taken before and after the science frame. The weighting of each sky frame is determined by the combination that gives the best uniform, zero background in the science frame. Prior to subtracting the sky frames from the science frames, twilight flats are normalized to remove solar spectra and then used to flatten the science and sky data. Finally, cosmic rays are masked.

The reduced data thus consist of a galaxy spectrum, an arc spectrum, and noise for each fiber, along with fiber coordinate information. Typically a small number of fibers for a given galaxy are contaminated by light from neibhboring stars or galaxies or other data problems; we remove these unusable fibers by hand. We identify these fibers in two ways: by comparing a map of fiber fluxes of our data to published images of each galaxy, and by examining the 1D radial light profile. Most contaminated fibers are obvious outliers to the light profile, but some relatively dim interlopers in very elliptical galaxies may still be present along the minor axis, so we also check the 2D fiber flux map (comparing with published images) by eye for additional contaminated fibers. In the case of very bright and/or extended contaminants, we look at the 2D map and remove an additional ring of fibers around those that are obviously contaminated. This step is designed to remove fibers that may only be contaminated at the $\lesssim 10\%$ level, and so would not show an obvious excess of flux in the light profile.

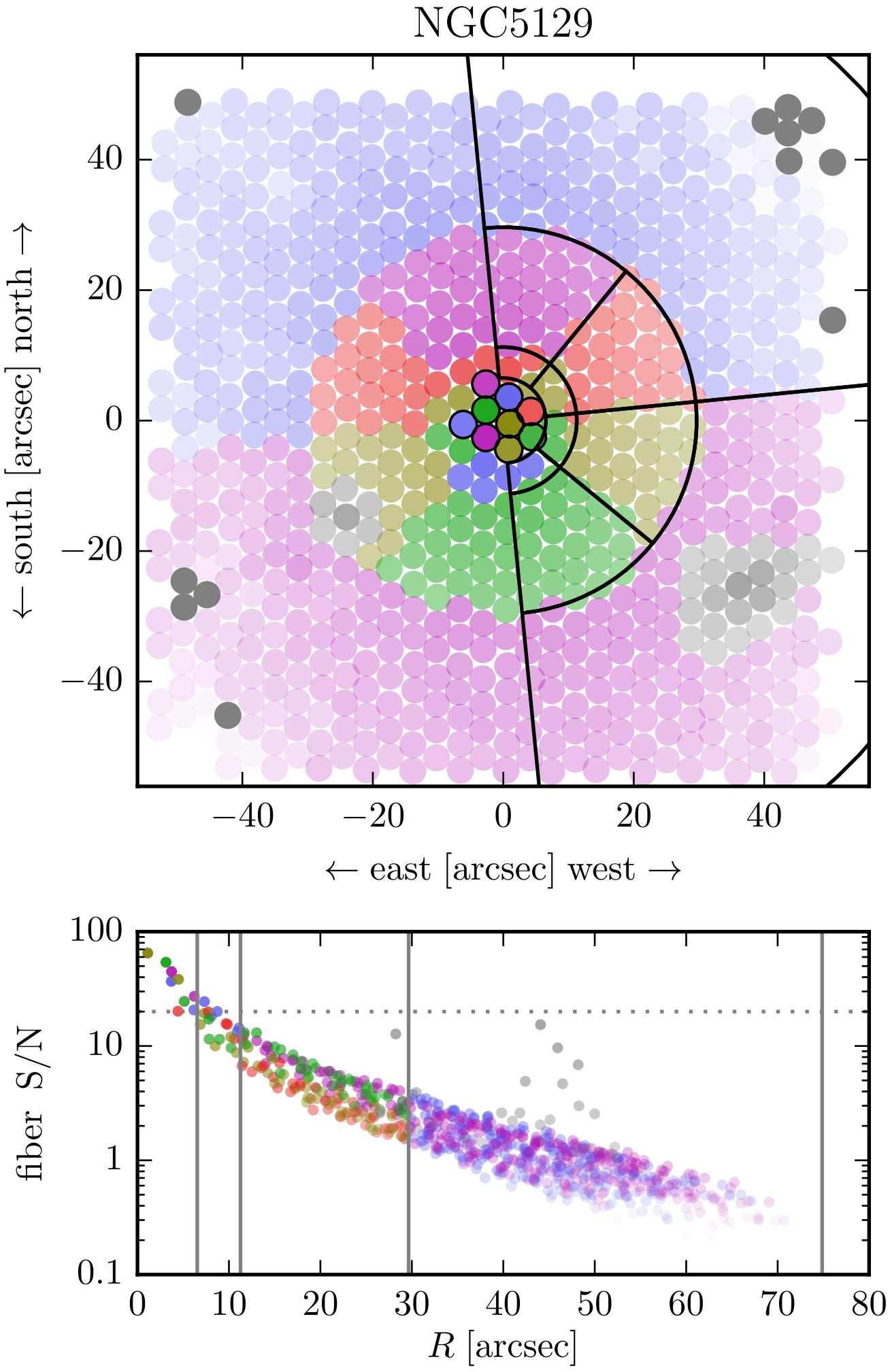

**Figure 1.** Example binning scheme for galaxy NGC 5129. Fibers from all three dithers are shown, providing a contiguous coverage of $\sim 100''$ by $100''$ field of view. The level of transparency corresponds to the total integrated flux from that fiber. Fibers in the same bin are shown with matching colors. The top panel shows the full field of view with bins outlined in black. The bottom panel shows the S/N of each fiber vs radius, with bin divisions shown as vertical lines and the S/N threshold shown as a dotted horizontal line. The central fibers each constitute a single bin, because their S/N already exceeds 20. Multi-fiber bins are folded across the major axis, with outlines shown only on one side in the top panel. A few fibers at the edges are discarded due to poor data quality (dark grey); some other fibers are excluded due to contamination from nearby objects (lighter gray).

## 3 KINEMATIC ANALYSIS

### 3.1 Spatial Bins

In the central regions of our galaxies, the spectra from individual fibers often have S/N exceeding 50, so we use these single-fiber spectra directly in the kinematic analysis. For the outer parts of the galaxies, we set a S/N threshold of 20, and any fibers with a lower S/N are combined into bins such that the resulting co-added spectrum reaches at least S/N = 20. More precisely, we find the largest radius within which all fibers have the required S/N of 20, and bin all fibers outside that radius into a circular binning scheme. An example is shown in Figure 1.

For the outer binned fibers, we first "fold" the fibers over the major axis and combine symmetrical bins to increase their S/N. The fibers are then divided into annular bins of varying radial size, and each annulus is cut into an even number of equal-sized angular bins. We require that the aspect ratio of each bin, defined as $[0.5(R_{\rm outer} + R_{\rm inner})\Delta\theta] / [R_{\rm outer} - R_{\rm inner}]$, not exceed 1.5, so the number of angular bins is effectively determined by the thickness of the annulus. Marching outward from the center, the thickness of each annulus is increased until the S/N of each folded bin in that annulus passes the minimum thresh-old. In many galaxies, some fibers at the outskirts of the field are discarded when the remaining outer fibers cannot achieve sufficient S/N. (This is not the case for Figure 1, where the outermost annulus contains many fibers and is very close to meeting the S/N threshold.) We use a luminosity-weighted average of the individual fiber radii to calculate the average radius of each bin.

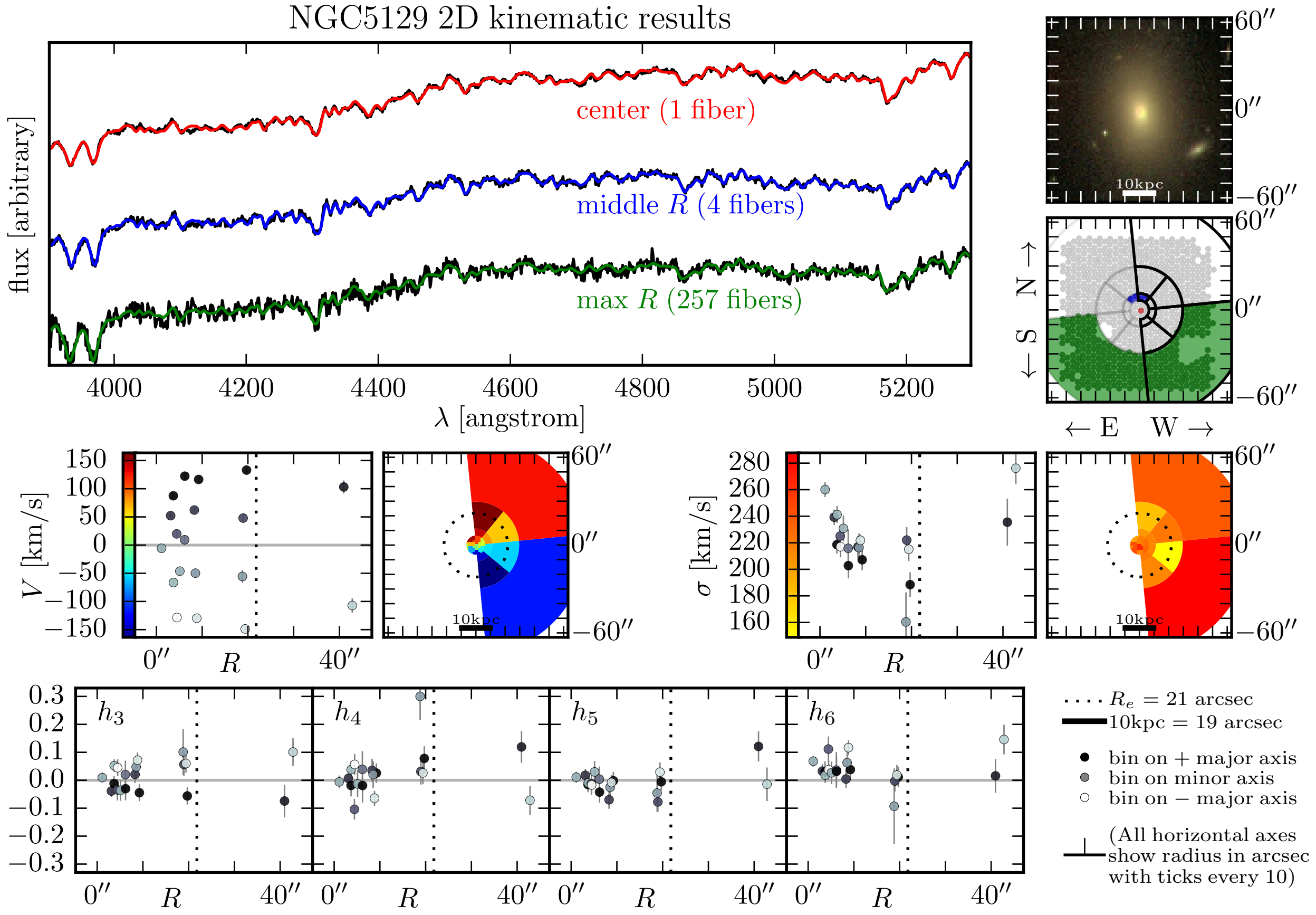

**Figure 2.** Example kinematic results for NGC 5129. Three representative spectra are shown (top row) for the center fiber, a bin at intermediate radius, and an outermost bin, respectively. The corresponding bin locations are shown on the right, along with the galaxy image. For each spectrum, the best-fit pPXF result (in color) is superimposed on top of the observed spectrum (black). The radial profiles of all six velocity moments are shown (middle and bottom rows). At each radius, the moments for the various azimuthal bins are shaded in decreasing grey scale from black (for bins along positive major axis), to grey (minor axis), to white (negative major axis). The 2D maps of the stellar velocity $V$ and dispersion $\sigma$ are also shown. $R_e$ is shown by a dotted line, and the physical scale is shown by a thick bar representing 10 kpc in the $V$ and $\sigma$ maps. A condensed version of this figure is provided for each galaxy in Appendix D.

### 3.2 Stellar templates and velocity distribution measurements

We use the penalized pixel-fitting (pPXF) method (Cappellari & Emsellem 2004) to extract the stellar line-of-sight velocity distribution (LOSVD) function, $f(v)$, from the absorption line features of each spectrum. This method convolves a set of spectra from template stars with $f(v)$ modeled as a Gauss-Hermite series up to order $n = 6$:

$$f(v) \propto \frac{e^{-\frac{(v-V)^2}{\sigma^2}}}{\sqrt{2\pi\sigma^2}} \left[1 + \sum_{m=3}^{n} h_m H_m \left(\frac{v-V}{\sigma}\right)\right], \quad (2)$$

where $V$ is the mean velocity, $\sigma$ is the dispersion, and $H_m(x)$ is the $m$th Hermite polynomial given by

$$H_m(x) = \frac{1}{\sqrt{m!}} e^{x^2} \left(-\frac{1}{\sqrt{2}} \frac{\partial}{\partial x}\right)^m e^{-x^2}. \quad (3)$$

The third moment $h_3$ is a measure of the skewness of the distribution and the fourth moment $h_4$ is a measure of kurtosis. Because we fit up to $n = 6$, we also have $h_5$ and $h_6$ as parameters to further refine the fit. Our initial guess is 0 for $V$ and $h_3$ through $h_6$, and 250 km s$^{-1}$ for $\sigma$. We run the fits without penalty (i.e. setting keyword BIAS to zero), which means deviations from a Gaussian solution are not penalized.

We model the stellar continuum with an additive polynomial of degree zero (i.e. an additive constant only) and a multiplicative polynomial of degree seven. These polynomials are added to (and multiplied by) the template spectrum before convolving with the LOSVD. The polynomial coefficients and the six velocity moments are fit simultaneously.

For the stellar templates, we use the MILES library of 985 stellar spectra (Sánchez-Blázquez et al. 2006; Falcón-Barroso et al. 2011) and run pPXF over the full library for each galaxy. This process typically returns nonzero weights for only about 20 of the template stars and is a time-intensive process, so we optimize over the full library only once for each galaxy, using the high-S/N full-galaxy spectrum from co-adding all fibers for that galaxy. To obtain the spatially-resolved $f(v)$ for multiple locations within a galaxy, we then fit each individual fiber (for a galaxy's inner region) or binned spectrum (for the outer region) with the $\sim 20$ stellar templates chosen in the full-galaxy fit, allowing pPXF to determine the best-fit template weights over this restricted list of available templates.

For the wavelength range of the fit, we crop each of our Mitchell spectra to a range of 3900 – 5300 Å. Any prominent emission lines are masked. The MILES library covers the wavelength range 3525-7500 Å at 2.5 Å (FWHM) spectral resolution. To account for the instrumental resolution of the Mitchell spectrograph, we convolve the stellar templates with a Gaussian distribution of an appropriate dispersion that is determined individually for each bin. The instrumental resolution varies by factors of about 20% over the wavelength range of the fit, but is typically around 4.5 Å FWHM (see details in Section 2.2).

We perform Monte Carlo calculations to determine the error bars on the best-fit velocity moments returned by pPXF. We define a noise scale using the actual noise of each spectrum, and add randomized Gaussian noise to each spectral pixel to create 100 trial spectra for each bin. The error for each moment is then the standard deviation of the pPXF fit results from the 100 trial spectra.

### 3.3 Example Kinematics and Comparisons to literature

Figure 2 shows an example of our kinematic results for NGC 5129. A condensed version of this figure is provided for all 41 galaxies in Appendix D, showing the first 4 velocity moments ($V$, $\sigma$, $h_3$, $h_4$) as well as the galaxy images and fiber and bin maps. Later sections discuss in detail our results and implications for each velocity moment: $V$ and angular momentum $\lambda$ in Section 4, $\sigma$ in Section 5, and $h_3$ through $h_6$ in Section 6. Our measurements of $\lambda$ and $\sigma$ within $R_e$ and the central $\sigma$ for all 41 galaxies are listed in Table 1.

For galaxies with existing kinematics in the literature, we find general good agreement. In Appendix A, we compare our central-fiber $\sigma_c$ with the values listed in Hyperleda and NSA. Six galaxies are in common between the MASSIVE and ATLAS$^{3D}$ surveys (Ma et al. 2014); among them, only NGC 4472 is in the high-mass subsample studied in this paper. Appendix B shows the excellent agreement between the kinematics from ATLAS$^{3D}$ and our results for NGC 4472, as well as NGC 5322 and NGC 5557. (The latter two galaxies are from our lower-mass sample and are included here for comparison purposes.) The MASSIVE data generally show less scatter at a given $R$ and cover two to five times farther in radius.

## 4 ANGULAR MOMENTUM

We measure the angular momentum of each galaxy using the dimensionless parameter $\lambda$ (Binney 2005; Emsellem et al. 2007)

$$\lambda(<R) \equiv \frac{\langle R|V|\rangle}{\langle R\sqrt{V^2+\sigma^2}\rangle} . \quad (4)$$

Averages here refer to luminosity-weighted averages of $R|V|$ or $R\sqrt{V^2+\sigma^2}$ over all spatial bins enclosed within radius $R$, indicated by the notation $\lambda(<R)$. We also measure a *local* $\lambda$, where the above average is calculated only over the bins in the same annulus at $R$, indicated by the notation $\lambda(R)$.

The parameter $\lambda$ is used in a similar way as $V/\sigma$ to quantify the dynamical importance of rotation relative to dispersion in a galaxy. While $\lambda$ contains both $V$ and $\sigma$ in its definition, for our sample it is mostly sensitive to the details of $V$ and not $\sigma$ (see Section 4.2).

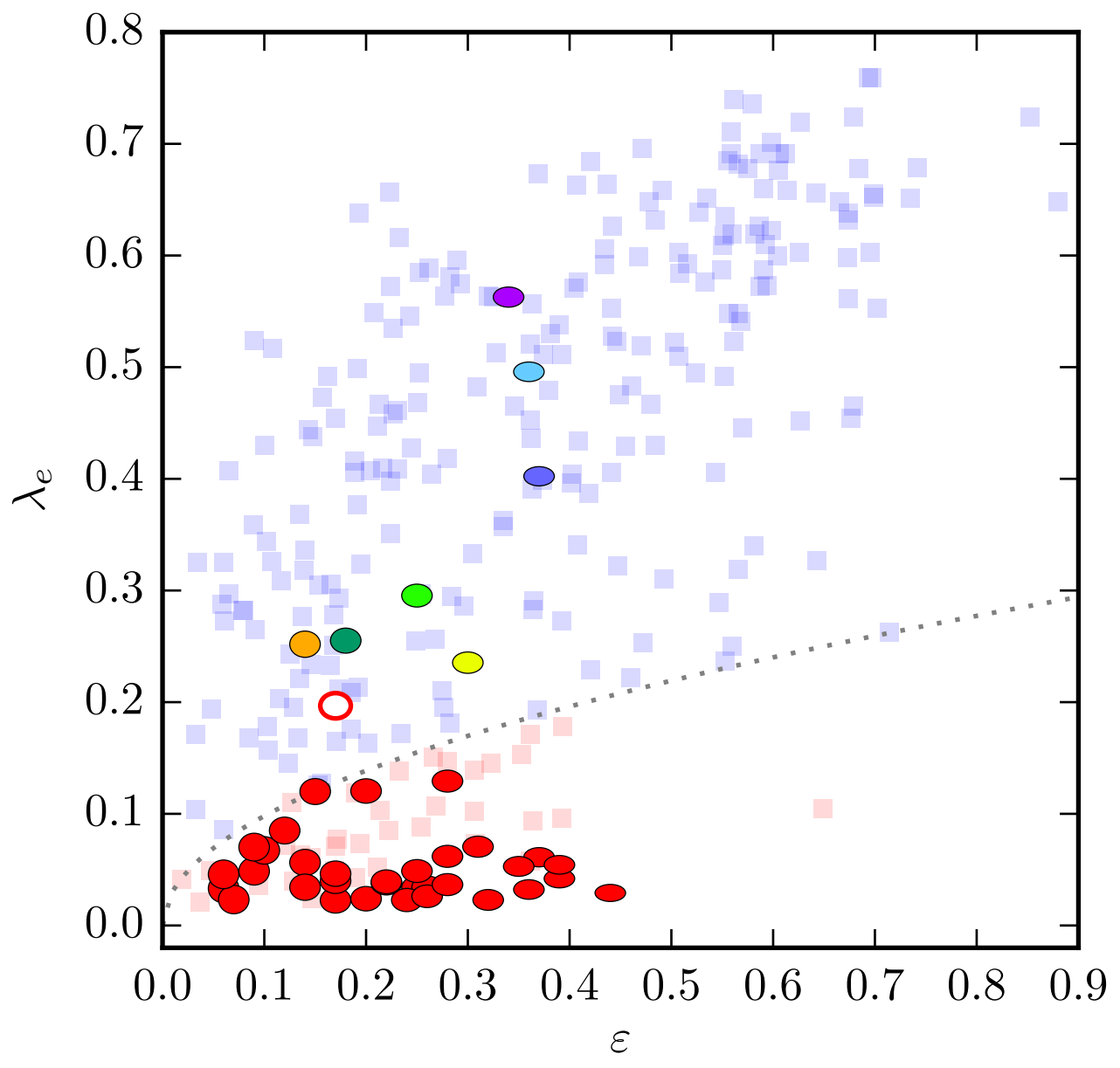

**Figure 3.** Angular momentum within $R_e$ ($\lambda_e \equiv \lambda(<R_e)$) vs ellipticity $\varepsilon$ for the 41 MASSIVE galaxies (circles/ellipses; symbol shape represents $\varepsilon$) and the ATLAS$^{3D}$ sample (faint squares). The gray dotted curve illustrates the cutoff of $\lambda_e = 0.31\sqrt{\varepsilon}$ between fast rotators and slow rotators used in ATLAS$^{3D}$ . All slow rotators are shown in red, while fast rotators are shown in blue for ATLAS$^{3D}$ galaxies and color-coded individually for MASSIVE galaxies (see Figure 5).

### 4.1 Global angular momentum $\lambda_e$

Figure 3 shows the angular momentum within the effective radius $R_e$, defined as $\lambda_e \equiv \lambda(<R_e)$, vs ellipticity $\varepsilon$ for the 41 MASSIVE galaxies (circles/ellipses) and the 260 ATLAS$^{3D}$ galaxies (squares). The values of $R_e$, $\varepsilon$ and $\lambda_e$ for each MASSIVE galaxy are listed in Table 1. The black dotted curve indicates the cutoff between slow (red symbols) and fast rotators (blue and non-red symbols), $\lambda_e = 0.31\sqrt{\varepsilon}$, found empirically for the ATLAS$^{3D}$ sample (Emsellem et al. 2011). This criterion takes into account inclinations and applies specifically to measurements within an aperture of $R_e$.

Seven of the 41 MASSIVE galaxies in Figure 3 have $\lambda_e > 0.2$, which we classify as fast rotators: NGC 383, NGC 2320, NGC 2693, NGC 3158, NGC 3805, NGC 5129, and NGC 7426. They are color-coded individually in Figure 3 and subsequent figures. The 33 slow rotators are shown as filled red circles/ellipses. (We leave NGC 4472 unclassified as discussed below.) We find a much higher fraction of slow rotators (33/41 ∼ 80%), compared to 36 out of 260 ATLAS$^{3D}$ galaxies (∼ 14%). Figure 4 shows the fraction of slow rotators in bins of $M_K$ for the MASSIVE subsample studied here and the ATLAS$^{3D}$ galaxies. The slow fraction increases dramatically from ∼ 10% at $M_K \sim -22$ mag to ∼ 90% at $M_K \sim -26$ mag. Within the ATLAS$^{3D}$ sample, the fraction of slow rotators stays low until their highest luminosity bin ($M_K \sim -25$ mag) in which the slow fraction rises to ∼ 40% (Emsellem et al. 2011). The MASSIVE data demonstrate that the critical range for ETGs to transition from being predominantly fast to predominantly slow rotators occurs at $M_K \sim -25.5$ mag.

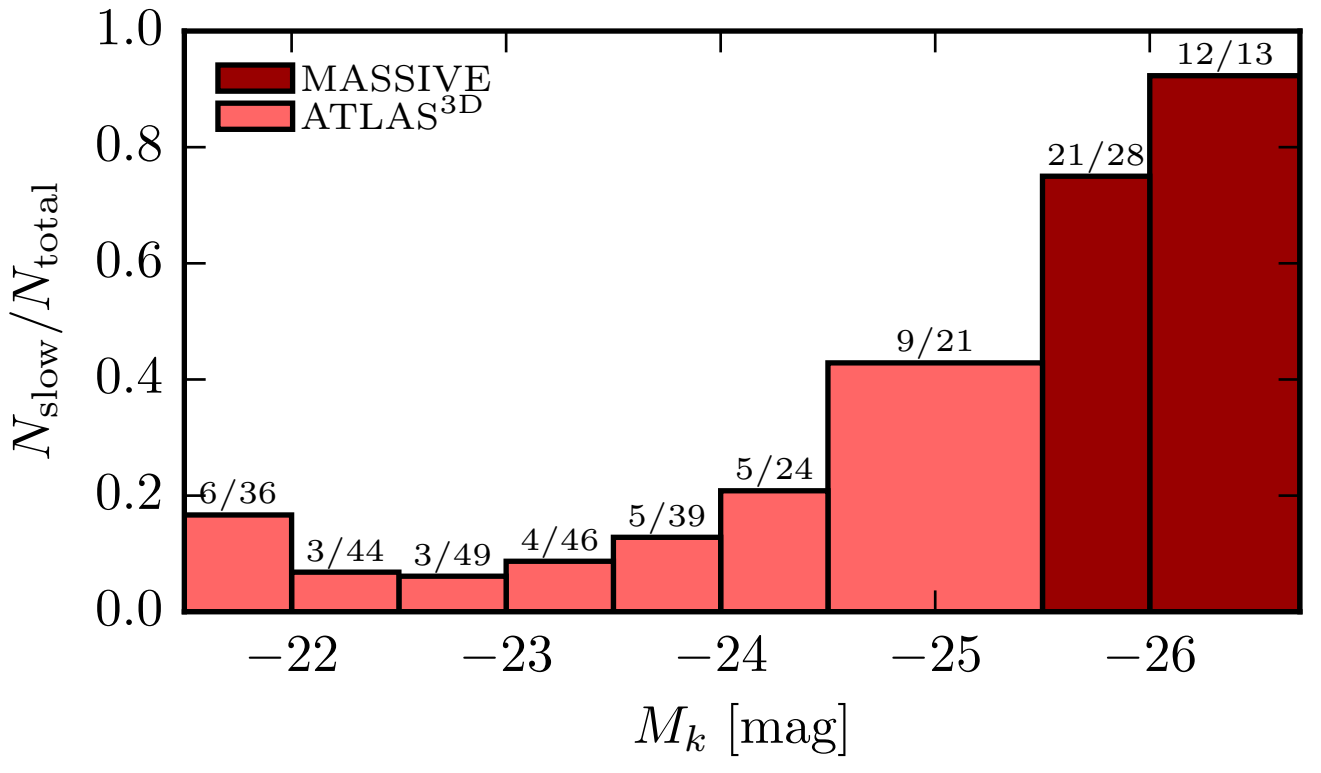

**Figure 4.** Fraction of slow rotators as a function of absolute $K$-band magnitude $M_K$ for the 41 MASSIVE galaxies and the ATLAS$^{3D}$ sample.

While the $R_e$ from NSA and 2MASS for our galaxies may be under-estimated (see Section 2), in practice, all of our galaxies (except NGC 4472) have nearly flat $\lambda(< R)$ profiles beyond $R_e$ (see top panel of Figure 5). Increasing $R_e$ therefore would not substantially impact the measurement of $\lambda_e$ or our classifications of fast vs slow rotators.

NGC 4472 is an example of the borderline cases for which the slow vs fast classication depends sensitively on the spatial extent within which the angular momentum is measured, and whether one accounts for $\varepsilon$ in the classification. NGC 4472 also happens to be the only galaxy in common between the 41 MASSIVE galaxies in this paper and the ATLAS$^{3D}$ sample. ATLAS$^{3D}$ reports $\lambda_e = 0.077$ and $\varepsilon = 0.172$ for NGC 4472 (using $R_e = 95.5''$) and classifies it as a slow rotator (Emsellem et al. 2011). Their IFS data, however, reach only a radius of $0.26R_e$ for this galaxy (or $0.14R_e$ for $R_e = 177''$). Our kinematics agree well with theirs out to this radius (see Appendix B and Fig. B1), but our large radial coverage of NGC 4472 shows that $V$ increases with radius[1] out to $\sim 160''$, and $\lambda$ increases from 0.13 at $R \approx 50''$ to $\approx 0.17$ at $R \approx 95.5''$ ($R_e$ used in ATLAS$^{3D}$ ), and flattens to $\approx 0.2$ at $R \approx 160''$ and beyond. In Table 1 and the rest of the paper, we therefore adopt the (circularized) $R_e = 177''$ from the 2D profile fits of Kormendy et al. (2009) and $\lambda_e = 0.2$ for NGC 4472. This value of $\lambda_e$ would result in classification as a fast rotator by the ATLAS$^{3D}$ criterion, but we note that our value may be biased slightly high due to the unusual pointing scheme: we took multiple pointings along the major axis to cover the large extent of this galaxy, resulting in more coverage along the major axis than minor axis (see Figure D8). The value of $\lambda_e$ is computed by averaging over all spatial bins, including those along the minor axis where $V$ and $\lambda$ are small. The "missing" minor axis coverage for NGC 4472 at the outer bins may therefore result in an inflated $\lambda$ at $50''$ and beyond. Furthermore, cored and non-cored elliptical galaxies separate quite cleanly below and above $\lambda_e \approx 0.25$ independent of $\varepsilon$ (Lauer 2012; Krajnović et al. 2013), and NGC 4472 falls among the cored slow rotators according to that classification. For all these reasons, we classify NGC 4472 as an intermediate case in this paper.

[1] Rising radial velocity along the major axis is also seen in several earlier long-slit datasets of this well-studied galaxy (e.g. Davies & Birkinshaw 1988; Franx et al. 1989; van der Marel et al. 1990; Bender et al. 1994)

### 4.2 Radial profiles of $\lambda$

Figure 5 shows the (cumulative) angular momentum $\lambda(< R)$ (top panel), local $\lambda(R)$ (middle), and velocity curves (bottom) for the 41 MASSIVE galaxies. The color-coding is identical to that in Figure 3, with slow rotators in red and fast rotators labeled individually. All three quantities are luminosity-weighted and follow similar overall radial shapes. The separation of the curves in the top panel into slow and fast rotator groups is preserved in the lower two panels, indicating that our assignment of slow vs fast rotators is reasonably robust and does not depend on the exact choice of parameter used to quantify galaxy rotations.

The similarity of the radial profiles for $\lambda$ vs $|V|$ also indicates that the shapes of $\lambda(< R)$ and $\lambda(R)$ are primarily driven by $V$ and not by $\sigma$, despite the fact that $\sigma \gg V$ in most bins for all galaxies in our sample. This is because $\sigma$ varies overall by a factor of 2 to 3 (individual bins ranging from about 150 to nearly 400 km/s), while $V$ varies by a factor of ∼ 10 (∼ 20 − 200 km/s).

For ease of comparison between $\lambda$ and $|V|$, we evaluate $|V|$ in the same way as $\lambda$ and plot in Figure 5 the radial profiles of the *average* of $|V|$ over all angular bins within each radial annulus. In comparison, long-slit observations typically measure the velocities along the major axis. The magnitude of the velocity shown here is reduced by averaging over bins far away from the major axis (or other rotation axis) and is likely to be smaller than the maximum rotational velocity, in particular for fast rotators. One exception is NGC 4472, as discussed in the previous subsection.

### 4.3 Gradients in $\lambda$ profiles

To further quantify the radial profiles of $\lambda$, we use the difference in local $\lambda(R)$ at two radii, $1.5R_e$ and $0.5R_e$, as a proxy for the gradient. In cases where the last annulus has a radius inside $1.5R_e$, we use the last data point. Similar analyses (with slightly different radii) are shown in Figure 11 of Arnold et al. (2014) and Figure 9 of Foster et al. (2016) for the 25 ETGs of the SLUGGS survey, and Figure 11 of Raskutti et al. (2014) for 33 massive ETGs.

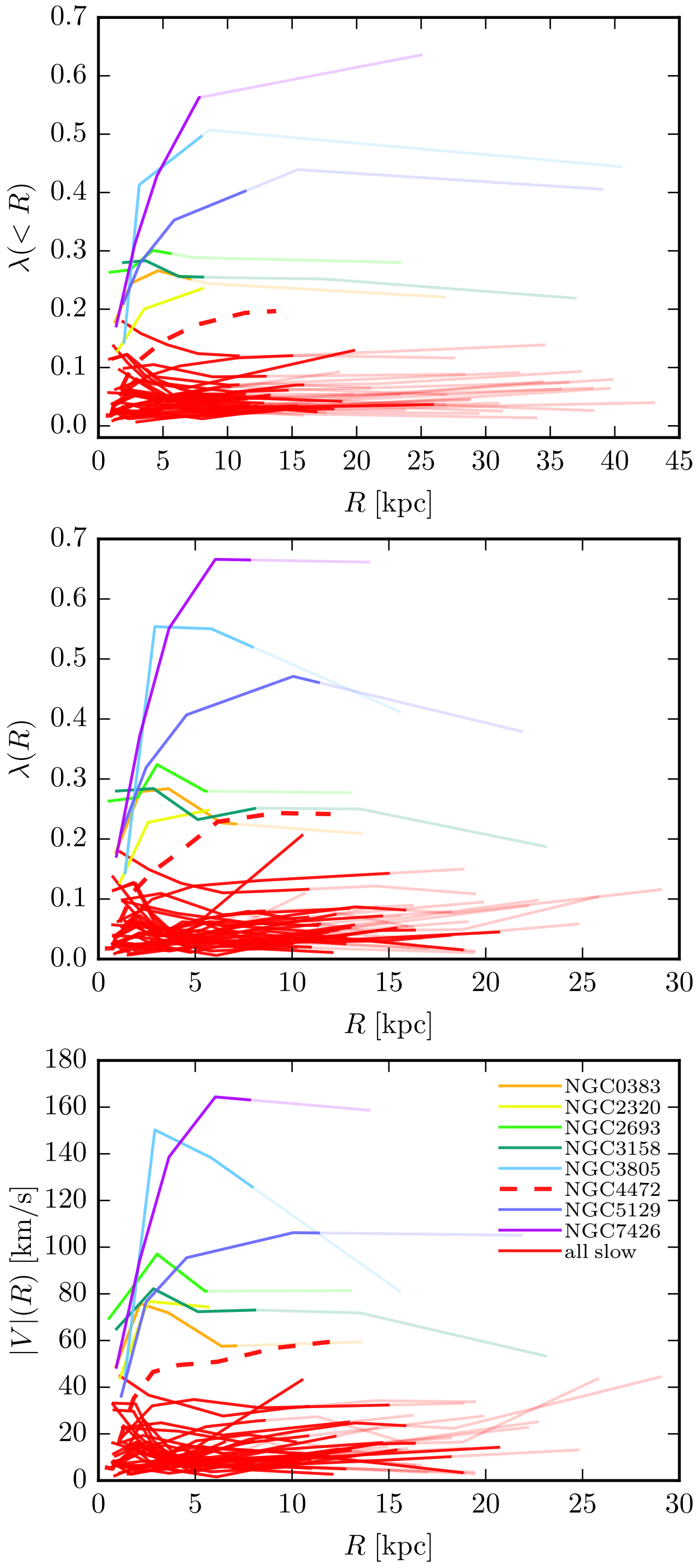

**Figure 5.** Radial profiles of angular momentum ($\lambda(< R)$ in top panel; *local* $\lambda(R)$ in middle panel) and velocity (bottom) for the 41 MASSIVE galaxies. All three quantities are luminosity-weighted, either within the annulus centered at $R$ (for $|V|$ and $\lambda(R)$) or over all bins/annuli enclosed by $R$ (for $\lambda(< R)$). All panels are color-coded in the same way, with slow rotators in red and fast rotators as listed in the bottom panel. The half-light radius $R_e$ is indicated by using fainter lines outside of $R_e$. Note the radial extent of each curve in the top panel goes out to the *maximum* extent of the last annulus, whereas the *center* of the last annulus determines the radial extent for the local $\lambda(R)$ and $|V|(R)$.

Figure 5 shows that all 7 fast rotators in our sample have flat or mildly declining $\lambda$ radial profiles[2]. Foster et al. (2016), on the other hand, finds most of the 25 SLUGGS galaxies to have a (mild) positive gradient in the local $\lambda$. Within our small sample of fast rotators, we do not find any significant trends between the $\lambda$ gradient and other galaxy properties. In comparison, Arnold et al. (2014) find their fast-rotating elliptical galaxies to have a negative gradient and S0 galaxies to have a positive gradient, whereas Raskutti et al. (2014) do not find such trends. We have only one fast-rotating S0 galaxy in our sample, NGC 383, and it has the most negative gradient. Further study is needed to assess whether the differences in $\lambda$ gradients between our sample and other samples (or among other samples) are due to small number statistics, differences in mass range or other sample properties, differences in gradient definition (e.g. using $\lambda(2R_e)-\lambda_e$), differing calibration of $R_e$ itself, or some other reason.

Most of our slow rotators in Figure 5 have quite flat $\lambda$ profiles beyond $\sim$ 5 kpc. Many have $\lambda < 0.05$ and undetectable rotational axes, consistent with being non-rotators (Ene et al. in prep).

### 4.4 Interesting velocity map features

In this subsection we highlight some notable features in the velocity maps in our sample. A more detailed analysis of the 2D velocity structures of MASSIVE galaxies such as kinematic twists, misalignments, and kinematically distinct cores (KDCs) and comparisons with other surveys (e.g., Krajnović et al. 2011; Fogarty et al. 2015; Foster et al. 2016). will be presented in a separate paper (Ene et al. in prep).

Two of our galaxies, NGC 1129 and NGC 4874, have clearly misaligned photometric and kinematic axes. NGC 1129 is misaligned by about 45 degrees, but the photometric axis is somewhat ambiguous since the ellipticity and PA both vary with radius. NGC 4874 is misaligned by nearly 90 degrees in the central region. Beyond the extent of our high S/N bins at $\sim 70''$, the photometric and kinematic axes of NGC 4874 appear reasonably well aligned.

Both NGC 1129 and NGC 4874 show signs of a twist in the kinematic major axis, which is not easily identifiable in our standard folded binning scheme. The twist is more apparent in an unfolded binning scheme with smaller spatial bins, which is being used for all galaxies in our ongoing study. The unfolded binning scheme will allow us to use kinemetry (Krajnović et al. 2006) to quantify the amount of twist of each galaxy. NGC 507 contains a clear KDC, a central fast-rotating component unconnected to the slow-rotating outer parts of the galaxy. Finding more subtle examples and quantifying these KDCs will also be possible with the unfolded binning and kinemetry analysis.

[2] A possible exception is NGC 7426, our fastest rotator, where $\lambda(< R)$ in our outermost bin is $\sim$ 15% higher than the neighboring bin. This slight rise is driven by the rapidly declining $\sigma$ in the outer part of this galaxy rather than $|V|$, which in fact declines slightly (bottom panel of Figure 5). The nominal gradient measured between $1.5R_e$ and $0.5R_e$ is much larger than other galaxies not due to this slight qualitative difference in $\lambda$ profiles, but because the inner point of $0.5R_e$ happens to fall on a lower point of the profile; a slight adjustment to $R_e$ might change the gradient calculation substantially, so we find it more instructive to look at the profiles qualitatively. Data beyond $r \sim 60''$ will reveal whether $\lambda$ continues to rise for NGC 7426.

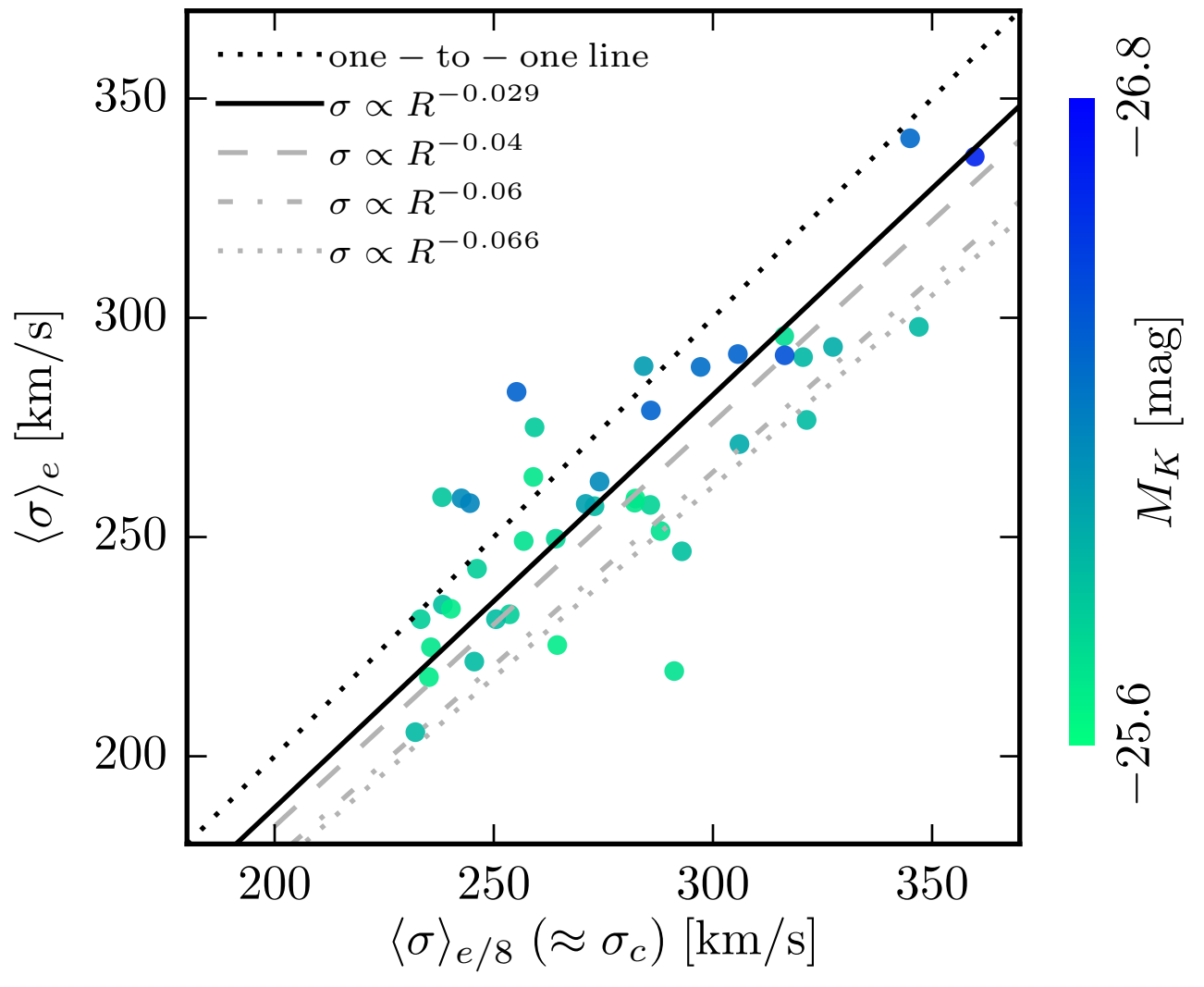

**Figure 6.** Comparison of velocity dispersion measured within two apertures, $R_e$ vs $R_e/8$, for the 41 MASSIVE galaxies. The latter is approximately the same as the velocity dispersion from our central fiber, $\sigma_c$, listed in Table 1. The solid black line shows the best-fit correction from this paper; the various gray lines show the aperture corrections using logarithmic slopes from the literature: −0.04 (Jorgensen et al. 1995), −0.06 (Mehlert et al. 2003), and −0.066 (Cappellari et al. 2006).

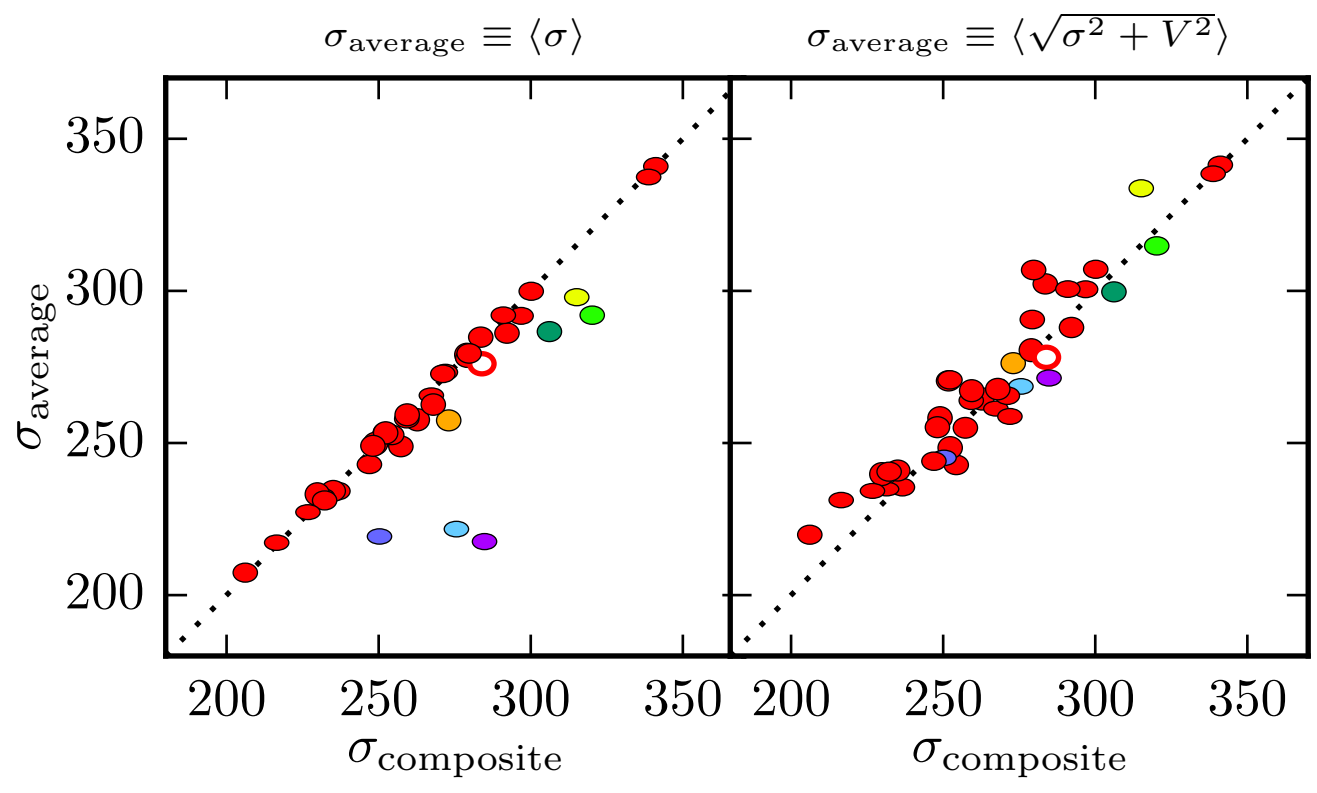

**Figure 7.** Velocity dispersion calculated from luminosity-weighted average over bins, compared to velocity dispersion calculated by fitting a single composite spectrum, for the 41 MASSIVE galaxies. Colors and shapes show fast/slow rotator status and ellipticity, as in Figure 3. In all cases, the $\sigma$ values shown here are calculated out to the same radius for both averages and composites (maximum $R$ ranges from 15 to 30″, adjusted to avoid asymmetries due to masked neighbors; this is very close to $R_e$ for many galaxies). The left panel shows the result of a simple luminosity-weighted average over $\sigma$ for each bin, which is in practice nearly identical to $\langle\sigma\rangle_e$ as listed in Table 1. The right panel shows the result of averaging over a combined $V$ and $\sigma$.

Finally, a small number of galaxies in our sample show systemic changes in velocity between the center and the outskirts of the galaxy. These include NGC 545, NGC 547, NGC 2256, NGC 2832, NGC 2892, and to a lesser extent, NGC 741 and NGC 1272. Most of these galaxies show an obvious visible companion (e.g., NGC 545 and NGC 547 are a close pair), suggesting that the outer parts of these galaxies may be slightly out of equilibrium. These galaxies are likely to have overestimated values of $\lambda$, because the definition of $\lambda$ does not distinguish between equal and opposite velocities on opposite sides of the galaxy and an overall systemic velocity shift. All of these galaxies are classified as slow rotators, with NGC 545 having the largest $\lambda_e$ value at 0.13. This puts NGC 545 near the boundary between fast and slow rotators, when in fact it shows no signs of rotation at all in the velocity map (see Figure D2). The other galaxies have $\lambda_e \lesssim 0.07$, which may still be overestimated but to a lesser degree.

## 5 VELOCITY DISPERSION

### 5.1 Global velocity dispersion $\langle\sigma\rangle_e$

To calculate a global value for velocity dispersion within an effective radius $R_e$ (denoted $\langle\sigma\rangle_e$), we take a luminosity-weighted average of $\sigma$ over all bins within $R_e$. The resulting values of $\langle\sigma\rangle_e$ for each galaxy in our sample is given in Table 1.

The spatially-resolved measurements of $\sigma$ enable us to compare $\sigma$ measured with different aperture sizes. Figure 6 compares $\langle\sigma\rangle_e$ with $\langle\sigma\rangle_{e/8}$, the luminosity-weighted average $\sigma$ within a radius of $R_e/8$. We find $\langle\sigma\rangle_{e/8}$ to be *smaller* than $\langle\sigma\rangle_e$ in 7 of the 41 galaxies; this small set of galaxies all have rising radial profiles $\sigma(R)$ (see next subsection). For the remaining $\sim 80\%$ of the galaxies, the central part of the galaxy dominates and $\langle\sigma\rangle_{e/8}$ is larger $\langle\sigma\rangle_e$.

We use the standard power-law form for aperture corrections and find the best-fit relation to be

$$\left(\frac{\langle\sigma\rangle_{e/8}}{\langle\sigma\rangle_e}\right) = \left(\frac{R_{e/8}}{R_e}\right)^{-0.029\pm0.036} . \quad (5)$$

This relation applies specifically to the correction between $R_e/8$ and $R_e$, as shown in Figure 6; because many of our $\sigma$ profiles are not well characterized by a single power law (see next subsection), corrections at different radii would have slightly different best-fit relations. The logarithmic slope of our relation is in reasonable agreement with the slope −0.04 of Jorgensen et al. (1995) used in the Hyperleda database, −0.06 from long-slit data for ETGs in the Coma cluster (Mehlert et al. 2003), and $-0.066 \pm 0.035$ for the SAURON sample (Cappellari et al. 2006). These various aperture corrections are shown as gray lines in Figure 6.

In Appendix A, we compare the velocity dispersion of our central fiber, $\sigma_c$, with the values of $\sigma$ across the literature. Even though our fiber diameter of 4″ does not cover a fixed fraction of $R_e$ for all galaxies in our sample, we find $\sigma_c$ and $\langle\sigma\rangle_{e/8}$ to be nearly identical.

We note that our method of measuring $\sigma$ within some aperture, by taking a luminosity-weighted average of $\sigma$ in each bin of our IFS data in that aperture, is *not* equivalent to measuring $\sigma$ from a single co-added spectrum within that aperture. The left panel of Figure 7 compares $\sigma$ from these two methods. It shows that the two $\sigma$ agree very well for slow rotators, but for fast rotators, $\sigma$ from a single co-added spec-trum is systematically higher than the luminosity-weighted $\sigma$ over IFS bins, in a few cases by as much as $\sim 30\%$. This difference is primarily a result of co-adding spectra over spatial bins with varying velocities $V$ (e.g., due to rotations). To assess the influence of $V$, we instead compare $\sigma$ from the co-added spectra to $v_{\rm rms} = \sqrt{\sigma^2 + V^2}$. The right panel of Figure 7 illustrates that $v_{\rm rms}$ gives a better approximation to $\sigma$ of the co-added spectrum for fast rotators, but it introduces scatter to the slow rotators.

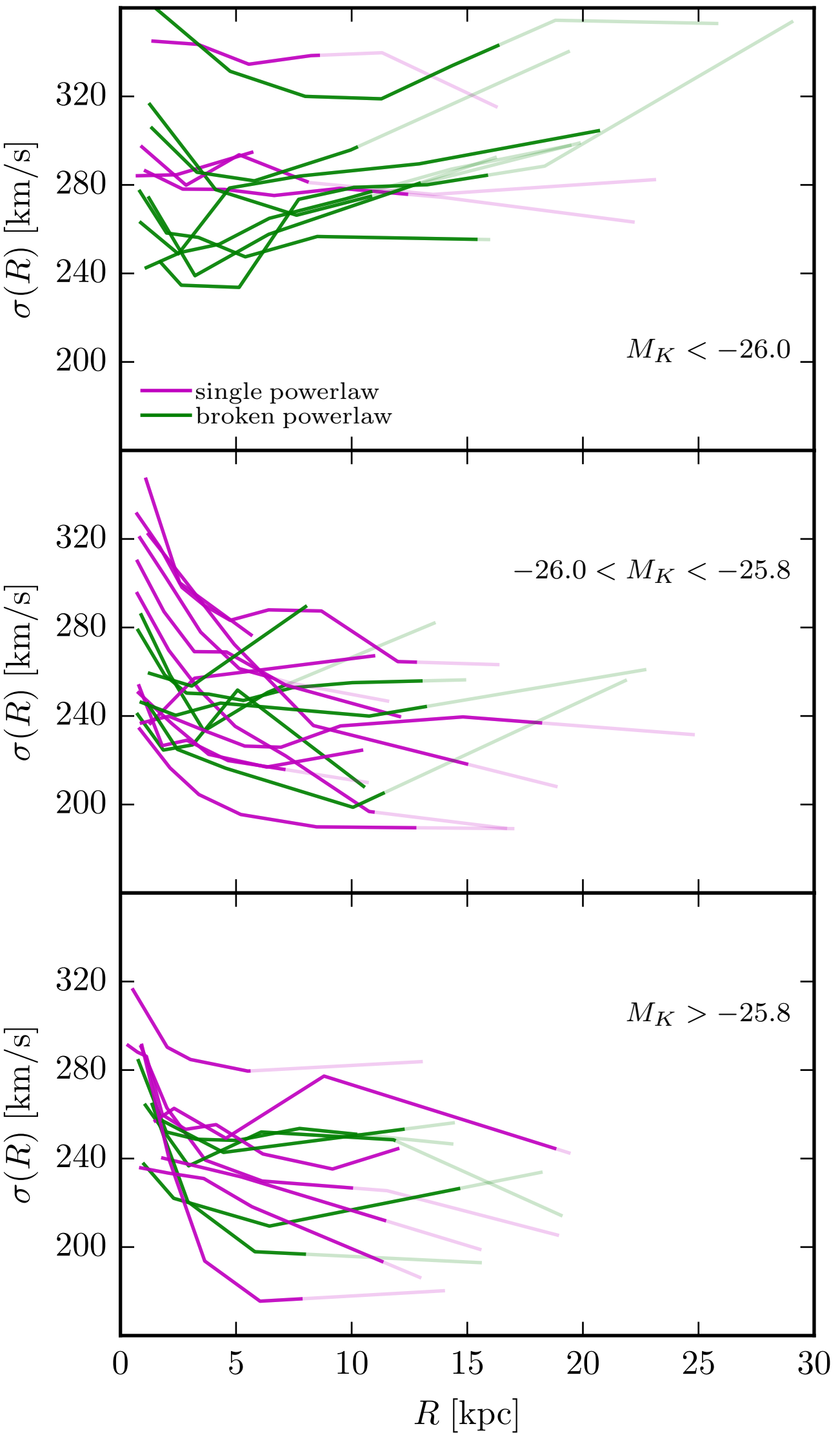

**Figure 8.** Radial profiles of velocity dispersion $\sigma$ for the 41 MASSIVE galaxies. The three panels show the sample in three $M_K$ bins, with roughly equal numbers of galaxies in each bin. The lines are color-coded by whether they are well-fit by a single power law (magenta) or a broken power law (green); see text for details. For each galaxy, the half-light radius $R_e$ is indicated by the line becoming fainter outside of $R_e$. Both the normalization and the shape changes for profiles in the different bins of $M_K$.

### 5.2 Radial $\sigma$ profiles, rising and falling

The radial profiles of the velocity dispersion $\sigma$ for the 41 MASSIVE galaxies are plotted in Figure 8, grouped into three bins by $M_K$. These profiles are analogous to the local profile $\lambda(R)$ in the sense that each point is a luminosity-weighted average over only the azimuthal bins in that annulus. The maximum $R$ here represents the "center" of the last annulus, not the total radial extent of the data.

The galaxies in the most luminous $M_K$ bin ($M_K < -26.0$ mag; top panel of Figure 8) have higher overall $\sigma$ as expected. The shape of the profile also changes with $M_K$: the most luminous galaxies all have flat or rising profiles, whereas the remaining galaxies (middle and bottom panels) show rising, flat, as well as falling profiles. This trend is in broad agreement with other observational studies that also find a diversity of $\sigma$ profiles with more rising profiles in more massive galaxies (Carter et al. 1999; Kelson et al. 2002; Loubser et al. 2008; Coccato et al. 2009; Pota et al. 2013; Murphy et al. 2014; Forbes et al. 2016), although none of these studies systematically probed galaxies as massive as in our survey.

To quantify the $\sigma$ profiles further, we fit to a broken power law:

$$\sigma(R) = \sigma_0\, 2^{\gamma_1-\gamma_2} \left(\frac{R}{R_b}\right)^{\gamma_1} \left(1 + \frac{R}{R_b}\right)^{\gamma_2-\gamma_1}, \quad (6)$$

where $\gamma_1$ is the power law slope at small radius, $\gamma_2$ is the power-law slope at large radius, and $R_b$ is the break radius. Due to degeneracies in the parameters, we fix $R_b$ to 5 kpc for all galaxies. (See Appendix C for details of the fit choices and parameter degeneracies.) About half of our galaxies are well fit by a single power law. All of these galaxies have $\gamma_1 = \gamma_2 \lesssim 0$, indicating that the $\sigma$ profiles either fall at all radii or remain nearly flat (magenta curves in Figure 8). For the remaining galaxies, a broken power law improves the $\chi^2$ per DOF of the fit by at least 0.3 compared to the single power law. We find these galaxies all to have a negative inner slope $\gamma_1$ and a positive outer slope $\gamma_2$, meaning the profiles decline at small radius from a central $\sigma$ peak but then flatten out or begin to rise at large radius (green curves in Figure 8). We do not find evidence of a bias in $\gamma_2$ from the radial extent of our observations.

Velocity dispersion is a central ingredient in dynamical modeling of the mass of ETGs, playing a role analogous to the velocity curves in disk galaxies. We will discuss implications of our dispersion profiles in this context in Section 7.

## 6 HIGHER ORDER GAUSS-HERMITE MOMENTS

### 6.1 Skewness $h_3$ and rotation

Figure 9 shows the (luminosity-weighted) average $h_3$ for each galaxy in our sample, vs $M_K$. The average $h_3$ for a galaxy is generally expected to be zero, and offsets from zero may indicate a template mismatch (Bender et al. 1994). Additional systematic offsets in $h_3$ arise when the galaxy centers do not align exactly with the center of a fiber. As a result, the most luminous (closest to center) fiber may have some nonzero $h_3$ due to being slightly off-center, especially in galaxies with strong central rotation. This effect would cause positive or negative offsets in $h_3$ randomly for each galaxy and contribute to some of the scatter in Figure 9. Overall, most galaxies in our sample have an average $h_3$ consistent with 0.

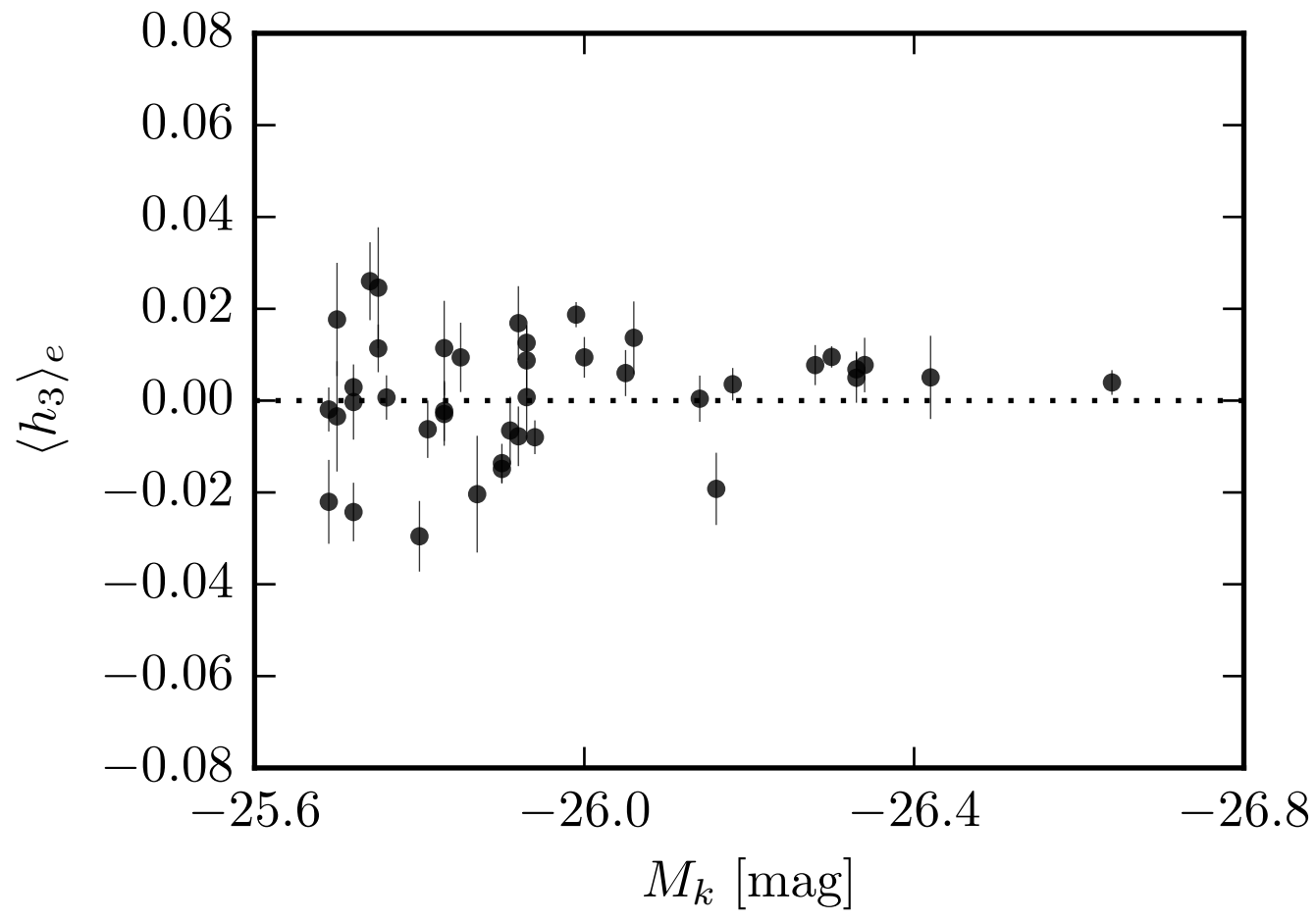

**Figure 9.** Average $h_3$ vs $M_K$ for the 41 MASSIVE galaxies in our sample. The value of $\langle h_3 \rangle_e$ is a luminosity-weighted average over all bins within $R_e$ of each galaxy.

The top two panels of Figure 10 show how the spatially-resolved $V$ and $h_3$ is anti-correlated within each of the seven fast rotators in our sample (coded by color). For each galaxy, the straight line shows our best fit to the relation between $h_3$ and $V/\sigma$. The slopes of this relation for all 41 galaxies are plotted against their angular momentum in the bottom panel of Figure 10. The anti-correlation between $V$ and $h_3$ is expected from projection effects. To illustrate this, consider the overall line of sight at any point along the major axis of an edge-on disk: it includes stars at the tangent point that contribute the highest $V$, as well as stars on larger orbits that are not at their tangent point. Together this creates a substantial "tail" of stars with smaller line-of-sight $V$, and skews the overall distribution. The approximate slope of this anti-correlation is −0.1 (Bender et al. 1994), which is indicated by the dashed horizontal line in the bottom panel of Figure 10. This panel shows that the $h_3$ vs $V/\sigma$ relation for our 3 most flattened and fastest rotators (NGC 7426, NGC 5129, NGC 3805) has a slope of almost exactly −0.1, while the more round and borderline fast rotators have somewhat steeper slopes. While observational studies have reported the general presence of this anti-correlation for rotating galaxies (e.g. Krajnović et al. 2011), and simulations have shown that gas can impact the relation (Hoffman et al. 2009), it would be interesting to examine in more detail how the trends in $h_3$ vs $V/\sigma$ slope emerge for larger samples of galaxies.

Numerical simulations of galaxy mergers have produced fast-rotating galaxies that lack a clear anti-correlation between $h_3$ and $V/\sigma$, called class D in Naab et al. (2014). In that work, five such galaxies are produced among a total of 44 simulated galaxies covering a mass range of $2\times10^{10}\ M_\odot \lesssim M^* \lesssim 6\times10^{11}\ M_\odot$. These five galaxies are formed in late gas-poor major mergers and have significant angular momentum without the signatures of embedded disk-like structures common to other fast rotators. Forbes et al. (2016) implemented the Naab et al. (2014) classifications and found only one tentative class D galaxy among the 24 SLUGGS galaxies studied. We have no class D galaxies among our 7 fast rotators, but we do not necessarily expect any for several reasons. The simulations were chosen to cover evenly a given halo mass range ($2.2 \times 10^{11} M_\odot \lesssim M_{\rm vir} \lesssim 3.7 \times 10^{13} M_\odot$) and not galaxy mass. The resulting $M^*$ of the simulated galaxies all lie below $M^*$ of the galaxies in our sample. In addition, our sample size of fast rotators is small, and baryonic physics in hydrodynamical simulations have well known uncertainties. Larger samples of both observed and simulated galaxies are needed to make any direct comparisons.

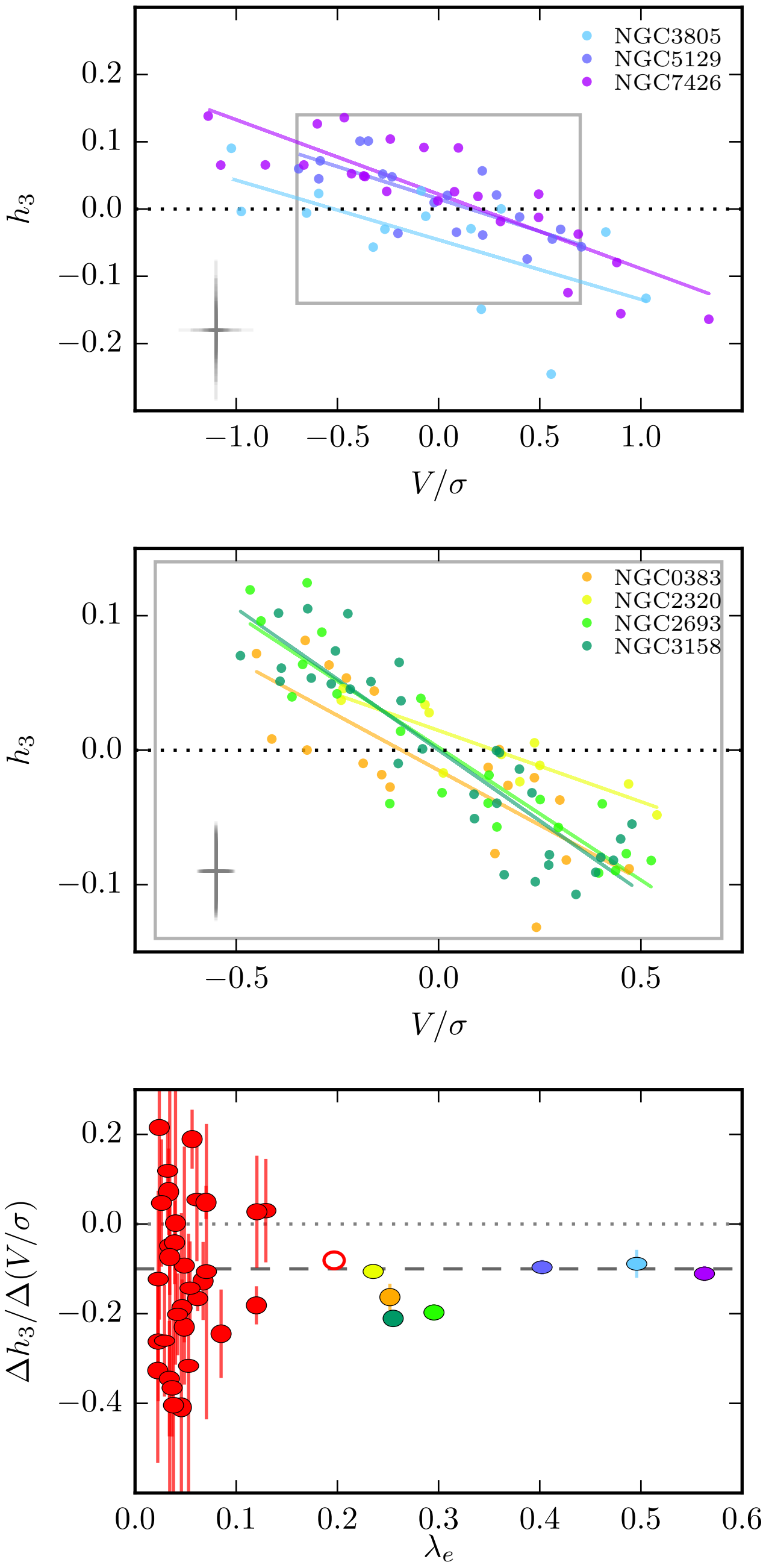

**Figure 10.** The two odd velocity moments, $V/\sigma$ and $h_3$, are anti-correlated within each of our fast-rotating galaxies. The straight lines in the top two panels show the best linear fit to $V/\sigma$ vs $h_3$ for the fast rotators, split into two panels to reduce crowding. Typical error bars for the data points are shown in the corners. The bottom panel illustrates the anti-correlations for our entire sample, plotting the best-fit *slope* in the top two panels vs $\lambda_e$, with the shape of the point corresponding to the real shape/flattening of the galaxy. The large error bars for the slow rotators (red symbols) in the bottom panel result from their narrow range of $V/\sigma$ and the uncertainties in determining the slopes. By contrast, the fast rotators with substantial flattening ($\varepsilon \gtrsim 0.3$, see also Figure 3) have a slope of −0.1, while the less flattened fast rotators have slightly steeper slopes.

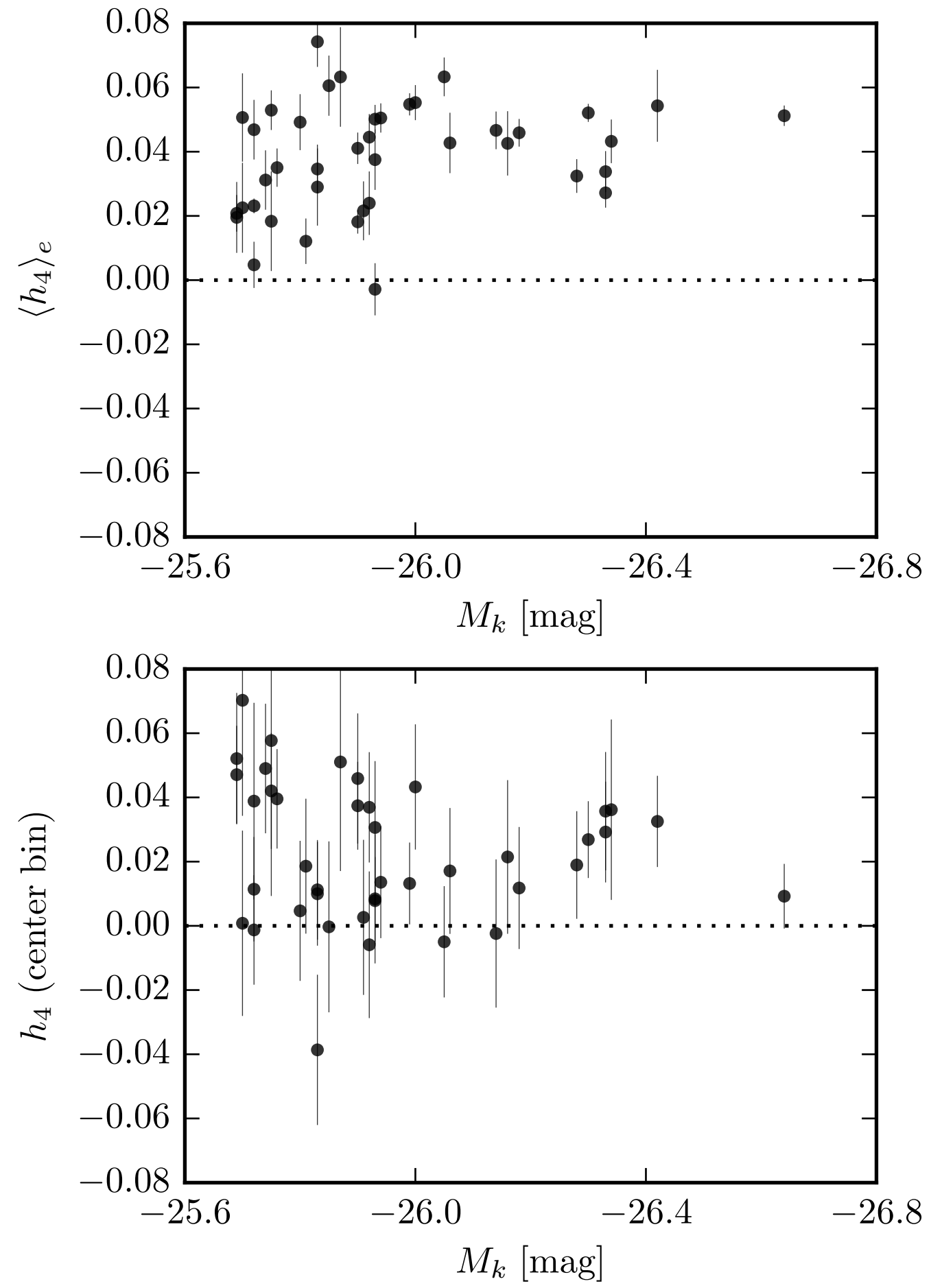

**Figure 11.** Average $h_4$ (top panel) and central $h_4$ (bottom panel) for each galaxy in our sample, vs $M_K$. The average $h_4$ is computed using a luminosity-weighted average over all bins within $R_e$ of each galaxy; the central $h_4$ is measured from the central fiber.

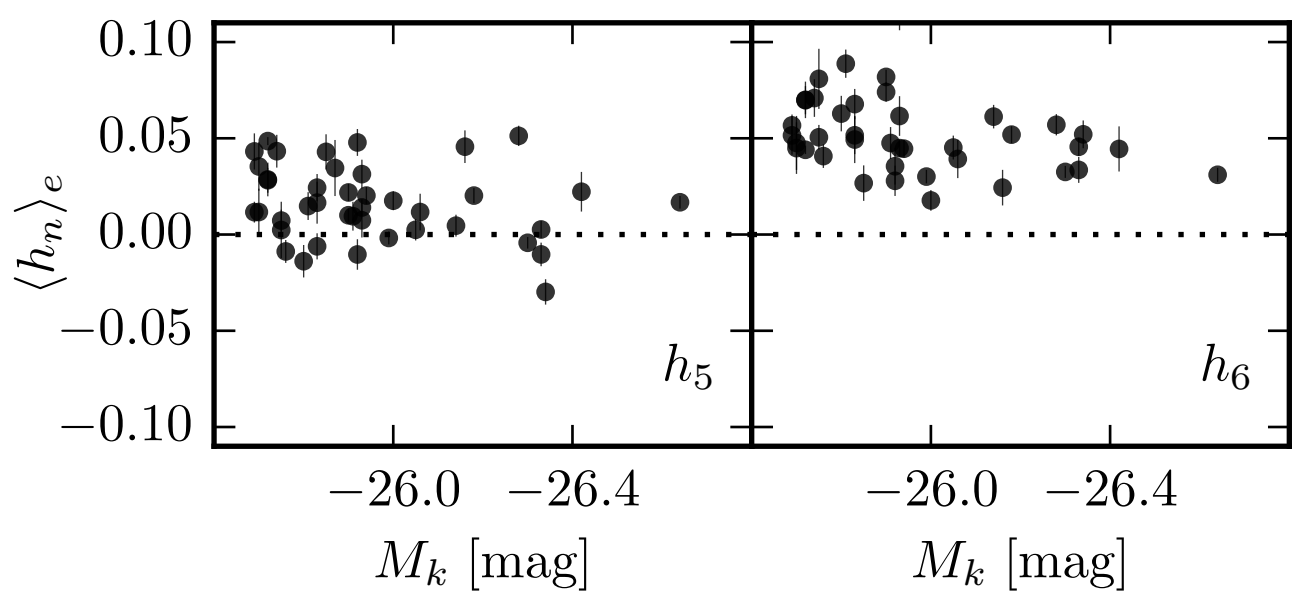

**Figure 12.** Average $h_5$ and $h_6$ vs $M_K$ for the 41 MASSIVE galaxies in our sample. Each point is computed using a luminosity-weighted average over all bins within $R_e$ for a given galaxy.

### 6.2 Kurtosis $h_4$

Figure 11 shows the fourth velocity moment $h_4$ vs $M_K$ for the 41 MASSIVE galaxies. The top panel shows the mean value $\langle h_4 \rangle_e$ over the galaxy (luminosity-weighted), and the bottom panel shows the central $h_4$ from the central fiber of each galaxy. Both $h_4$ are either consistent with 0 or positive overall. For many galaxies, $\langle h_4 \rangle_e$ is higher than the central $h_4$, indicating a prevalence of positive radial gradients in $h_4$.

The top panel of Figure 11 shows a clear trend for more luminous galaxies to have more positive $\langle h_4 \rangle_e$. Although not shown here, a plot with $\sigma_c$ on the $x$-axis instead of $M_K$ is qualitatively similar to Figure 11, as expected due to the correlation between $M_K$ and $\sigma$. We will discuss the possible implications of $h_4$, including the radial gradients, in Section 7.

### 6.3 Next-order deviations $h_5$ and $h_6$

The average $h_5$ and $h_6$ for the 41 MASSIVE galaxies are plotted in Figure 12. These higher moments of the LOSVD require high S/N spectra and are rarely measured. The left panel shows that $h_5$ is approximately centered around zero, similar to $h_3$, as expected for odd moments of the LOSVD. The right panel shows that $h_6$ is positive for all of our galaxies, somewhat mirroring $h_4$.

## 7 IMPLICATIONS FOR DYNAMICAL MASS AND ORBIT STRUCTURE

The line-of-sight stellar velocity dispersion $\sigma$ is a standard measure of the gravitational potential of a galaxy and is frequently used to infer a galaxy's dynamical mass. For a given measurement of $\sigma$, however, there is a well-known degeneracy between mass and velocity anisotropy (see, e.g., Binney & Mamon 1982; Gerhard et al. 1998; Thomas et al. 2007). A low line-of-sight $\sigma$ can be explained by *either* a low enclosed mass, *or* a radial velocity anisotropy that causes the true 3D velocity dispersion and hence the enclosed mass to be higher. A falling $\sigma$ profile with radius therefore does not necessarily imply the absence of massive dark matter halos, as illustrated by a series of papers with conflicting conclusions on the intermediate-mass elliptical galaxy NGC 3379 (Romanowsky et al. 2003; Dekel et al. 2005; Douglas et al. 2007; de Lorenzi et al. 2009).

The degeneracy in the line-of-sight $\sigma$ and velocity anisotropy for mass modeling can be alleviated somewhat by robust measurements of $h_4$, as we discuss later in this section.

### 7.1 Mass and $\sigma$

If $\sigma$ traces mass directly, then galaxies in large clusters or groups should see a rise in $\sigma$ at large $R$, increasing towards the cluster or group velocity dispersion. An apparent example is NGC 6166, in which the velocity dispersion is observed to rise from galaxy to cluster scale at large radius, reaching $\sigma \sim 800$ km/s (Bender et al. 2015).

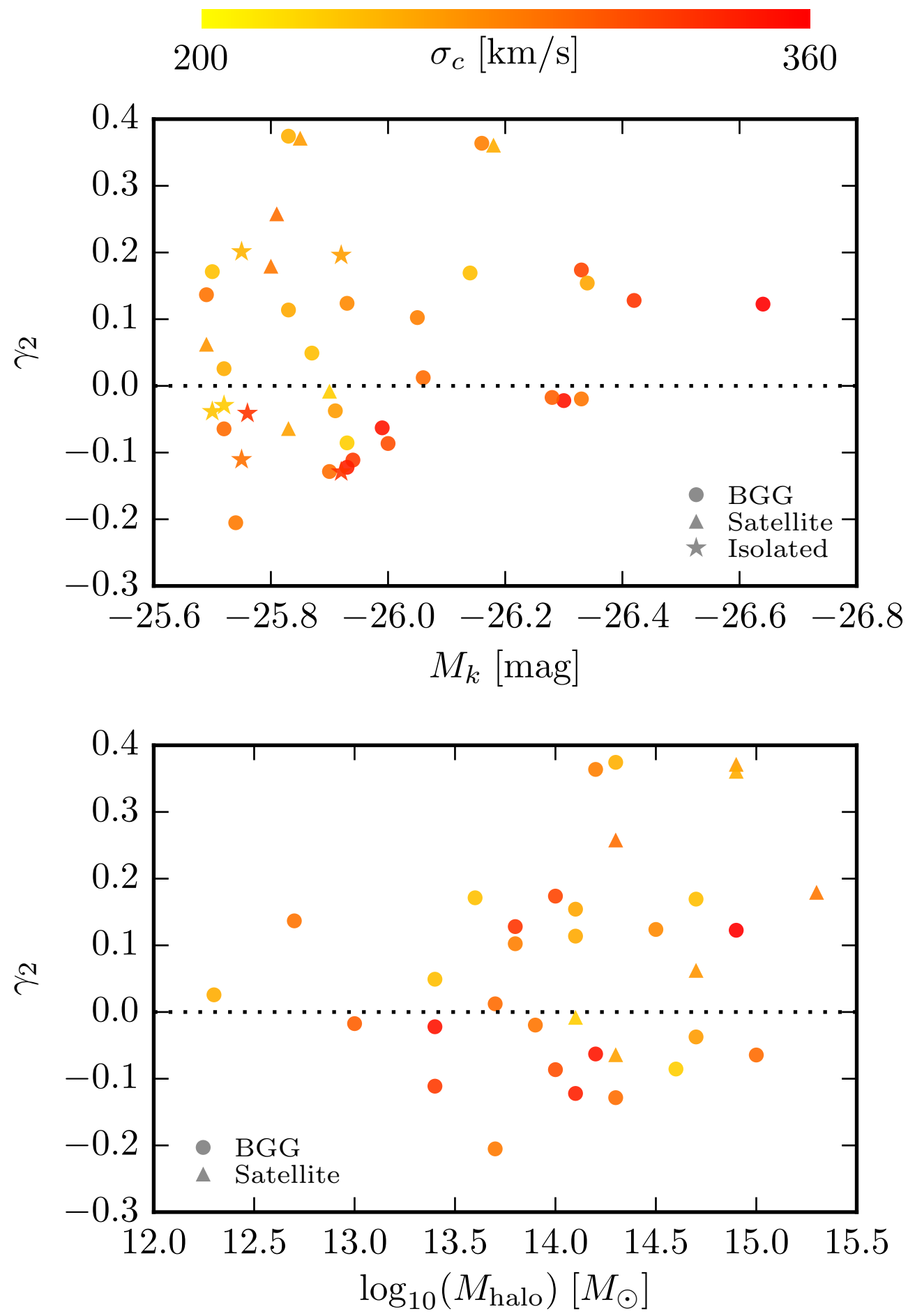

**Figure 13.** Large-radius behavior of $\sigma$ profiles (quantified by power law index $\gamma_2$, where $\gamma_2 > 0$ indicates a rising profile) vs galaxy luminosity $M_K$ (top) and dark matter halo mass $M_{\rm halo}$ (bottom) for the 41 MASSIVE galaxies. The three symbols distinguish three galaxy environments: brightest group galaxy (circles), satellite galaxy in a group (triangles), and "isolated" galaxy with fewer than 3 members in the 2MRS group catalog (stars). The central velocity dispersion $\sigma_c$ is indicated by color. At a given $M_K$, the highest values of $\sigma_c$ are associated with the lowest values of $\gamma_2$. Six out of the seven isolated galaxies have $\gamma_2 \lesssim 0$ (see top panel). Isolated galaxies have no halo mass measurement and are not shown in the bottom panel.

Figure 13 shows the slopes of the outer $\sigma$ profiles, $\gamma_2$, vs $M_K$ (top panel) and halo mass $M_{\rm halo}$ (bottom) for the 41 MASSIVE galaxies. Halo mass is taken from the virial mass estimator of the 2MRS HDC catalog (see Section 2.1). Different symbols indicate three larger-scale environments inhabited by MASSIVE galaxies: brightest group galaxies (BGGs) as circles, satellite galaxies in a group as triangles, and isolated galaxies with fewer than 3 members in the 2MRS group catalog as stars. As in Figure 8, we again see in the top panel that the 12 most luminous galaxies ($M_K \lesssim -26.0$ mag) all have nearly flat or rising profiles, i.e., $\gamma_2 \gtrsim 0$. Furthermore, all 12 galaxies except NGC 4874 are BGGs. By contrast, 5 out of the 7 isolated galaxies show falling outer $\sigma$ profiles ($\gamma_2 < 0$). These trends appear consistent with the presence of a larger group or cluster dark matter halo surrounding some BGGs. We note that not all BGGs have $\gamma_2 > 0$. As discussed above, declining $\sigma$ does not necessarily imply the absence of a dark matter halo.

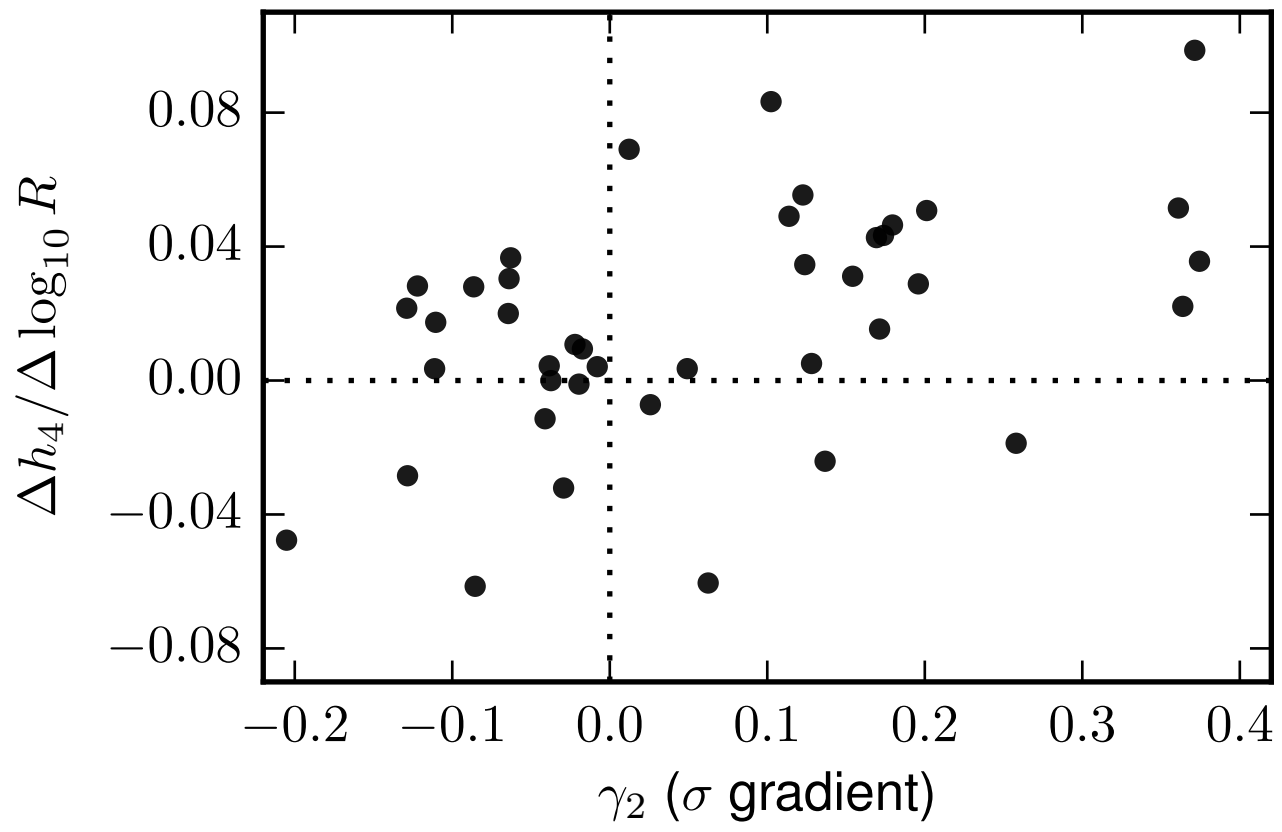

**Figure 14.** Radial gradients of $h_4$ vs $\gamma_2$, the gradient in velocity dispersion at large radius, as discussed in Section 5.2 and Appendix C. The two are positively correlated. In dynamical modeling, $h_4$ and $\sigma$ together are used to constrain both the mass profile and the velocity anisotropy; see text for discussion.

The bottom panel of Figure 13 shows a similar, albeit weaker, trend for galaxies in more massive halos to have a rising $\sigma$ profile. The scatter in this trend is large. Several galaxies with massive halos have falling profiles, and likewise several galaxies in smaller halos have rising profiles, including the two smallest halos. Galaxy environment can be characterized in a number of other ways beyond those shown here, and we will explore in more detail which (if any) environment measure correlates most closely with $\sigma$ profile behavior in a future paper.

### 7.2 Mass profiles and $h_4$

Before discussing how $h_4$ can help break the degeneracies among $\sigma$, mass, and velocity anisotropy, it is important to understand how $h_4$ behaves in *isotropic* systems. We also emphasize that the current discussion focuses on $h_4$, $\sigma$, and anisotropy behavior in the outskirts of the galaxy, with the goal of constraining the dark matter halo mass. Very similar models and arguments can be, and are, used to constrain the mass of central black holes (e.g. Thomas et al. 2016, and many others), but specific statements about the behavior of $h_4$ and other quantities that apply at the *center* of the galaxy may not apply to our discussion, and vice versa.

While $\sigma$ traces the circular velocity in isotropic systems, the exact shape of the LOSVD is important for disentangling mass and velocity anisotropy effects in galaxies where the orbit distribution is unknown. Analytic studies of spherical isotropic systems indicate that whenever the line-of-sight cuts through the galaxy regions with significantly different circular velocities (i.e. when the mass profile is not isothermal), then the LOSVD develops a core-wing structure with positive $h_4$ (e.g. Gerhard 1993; Baes et al. 2005). The more light coming from regions with a different circular velocity, the stronger this effect will be. Qualitatively speaking, this means that a strong increase (or decrease) in $\sigma$ with radius would be expected to cause an increase in $h_4$. The fact that our galaxies all have positive $\langle h_4 \rangle_e$ (Figure 11) can thus be explained by gradients in the circular velocity, without invoking velocity anisotropy. (This does *not* constitute evidence that there is no anisotropy in our galaxies, only that anisotropy is not necessary to explain this particular data feature.)

The connection between $h_4$ and mass profile shape may also be related to the positive correlations between the $h_4$ gradients and $\sigma$ gradients (Figure 14). There is no such correlation between $\langle h_4 \rangle_e$ and $\gamma_2$, although the above arguments might lead one to expect this correlation as well. As we will explain in the next section, it seems unlikely that velocity anisotropy would cause the correlation seen in Figure 14, so we speculate that the influence of gradients in circular velocity is the more likely cause.

### 7.3 Velocity anisotropy and $h_4$

Velocity anisotropy can add to the effects from the previous section and further influence $h_4$: at large radius, increased $h_4$ is associated with radial velocity anisotropy, and decreased $h_4$ with tangential anisotropy (e.g. Gerhard et al. 1998; Dekel et al. 2005; Douglas et al. 2007; Thomas et al. 2007). Physically, radial (or tangential) anisotropy can be thought of as an overabundance of stars at zero (or large) projected velocity causing a peaky (or boxy) shape to the LOSVD. Radial anisotropy also causes the projected (line-of-sight) dispersion to be an *under*-estimate of the three-dimensional dispersion, meaning that $\sigma$ will be suppressed; tangential anisotropy has the opposite effect, resulting in larger line-of-sight $\sigma$.

In the previous section we mentioned that positive $h_4$ may be caused by gradients in circular velocity, but radial anisotropy may also be a contributing factor to both the overall positive $\langle h_4 \rangle_e$ and to the trend we see with $M_K$ in Figure 11. Simulations have found that the details of merger conditions (e.g. spin alignment, impact parameter) can have a substantial effect on the anisotropy of the resulting galaxy (Dekel et al. 2005), and that a higher fraction of stars accreted from mergers (ex-situ formation) is connected to greater radial anisotropy (Wu et al. 2014). Combined with the finding in Rodriguez-Gomez et al. (2016) that higher mass galaxies tend to have a larger fraction of accreted (ex-situ) stars, this may explain why our more massive galaxies have more uniformly positive $h_4$.

In summary, the fact that many of our galaxies with increasing $\sigma$ at large radius also have positive and increasing $h_4$ suggests that they are unlikely to have isothermal mass profiles. This fact, however, does not provide strong constraints on the velocity anisotropy. If the mass profiles were isothermal, then for isotropic orbits we would expect a flat dispersion and $h_4 = 0$. While tangential anisotropy could increase the outer $\sigma$, it would make $h_4$ negative. Conversely, radial anisotropy could explain the observed positive $h_4$ but would cause a relative decline in $\sigma$ at large radii. For non-isothermal mass profiles, the rising $\sigma$ and rising positive $h_4$ can be attributed to the gradients in circular velocity, while still accommodating some range of velocity anisotropy that may cause secondary effects in $\sigma$ and $h_4$.

Gravitational lensing studies have found lensing ETGs to have a range of *total* mass profiles, from being nearly isothermal (e.g. Treu et al. 2006; Koopmans et al. 2009; Auger et al. 2009; Sonnenfeld et al. 2013), to having shallower profiles of a mean logarithmic density slope $-1.16$ (Newman et al. 2013). Most ETGs in the former studies are below the mass range $M_* \gtrsim 10^{11.8}$ $M_\odot$ studied here, whereas Newman et al. (2013) specifically targeted BCGs in massive, relaxed galaxy clusters of virial mass $\sim 10^{15} M_\odot$. A sample of 10 ETG lenses on galaxy-group scales suggests possible steepening in inner mass profiles with decreasing halo mass (Newman et al. 2015). Axisymmetric dynamical modeling based on the Jeans equation finds a sample of 14 ATLAS$^{3D}$ fast rotators to be well described by nearly isothermal profiles (Cappellari et al. 2015). Our ongoing dynamical mass modeling analysis will uncover the mass and velocity anisotropy profiles of MASSIVE galaxies.

### 7.4 Further Analysis

To properly disentangle the connections discussed in this section among mass, circular velocity, $\sigma$, anisotropy, and $h_4$ profiles, more detailed dynamical modeling is needed. These degeneracies can be resolved better with data that extends well beyond the radius where the model aims to constrain the galaxy properties (as emphasized in Morganti & Gerhard 2012), so the large radial extent of our data would be an important advantage in this regard. Some features of our data are also not fully captured by simple constant power-law measures such as $\gamma_1, \gamma_2$ or $\Delta h_4 / \Delta \log_{10} R$, and these features can be better leveraged by more direct modeling. For example, we see some signs that $h_4$ profiles with positive gradients tend to flatten out or begin declining at large radius (see Appendix D for $h_4$ profiles of each galaxy). This is in line with Bender et al. (2015), which found that $h_4$ rose until about 50 arcsec in NGC 6166 and then turned over, behavior that we have not attempted to capture in the current analysis. More detailed comparisons with simulations, e.g. similar to Remus et al. (2013), Wu et al. (2014), and Naab et al. (2014) but with additional focus on $\sigma$ at large radius, would also be useful.

## 8 SUMMARY

In this paper we presented the stellar kinematics of the 41 most massive galaxies ($M^* \gtrsim 10^{11.8}$ $M_\odot$) in the MASSIVE survey, a volume-limited sample of the highest end of the galaxy mass function. We reported the 2D kinematic measurements out to 1 to 4 times the effective radius of each galaxy from the Mitchell IFS, and discussed implications for the structure and evolution of these massive galaxies. Our high S/N IFS data enabled us to measure the 2D spatial distributions of the six Gauss-Hermite moments of the LOSVD ($V$, $\sigma$, $h_3$, $h_4$, $h_5$, and $h_6$), providing a rich dataset for future detailed modeling.

For each galaxy, we measured the radial profiles of the angular momentum parameter $\lambda$ and found our sample to have the following properties:

- More massive galaxies tend to have a larger fraction of slow rotators. We have 7/41 fast rotators, compared with

224/260 in ATLAS$^{3D}$ . We find the fraction of slow-rotators to increase sharply with galaxy mass, reaching $\sim$ 50% at $M_K \sim -25.5$ mag and $\sim$ 90% at $M_K \lesssim -26$ mag. (Figure 3, Figure 4)

- Most fast rotating galaxies show a moderately negative gradient in $\lambda$. There are no apparent trends between $\lambda$ gradient and morphology or other properties for our fast rotators, although our small number statistics make it impossible to draw strong conclusions. (Figure 5)
- Each of our fast rotators shows a clear anti-correlation between $h_3$ and $V/\sigma$ within the galaxy, as expected for galaxies with embedded disk-like components. We fit the *slope* of the anti-correlation between $h_3$ and $V/\sigma$ for each galaxy individually, and find a separation between faster, more flattened rotators and borderline, less flattened rotators: the 3 fastest rotators ($\lambda_e \gtrsim 0.4$ and $\varepsilon \gtrsim 0.3$) all show slopes of almost exactly $-0.1$, while more round galaxies ($\lambda_e \sim 0.2$ to 0.3, $\varepsilon \sim 0.1$ to 0.3) show steeper slopes up to $-0.2$. (Figure 10)

We also investigated the radial profiles of $\sigma$, and found the following properties:

- The radial $\sigma$ profiles show diverse shapes at both small and large radius. We quantify this by fitting a power law to the profiles, and find roughly half of the galaxies require a broken power law (where we fix the break radius to 5 kpc) to accommodate a shape that initially falls from a high central value but then turns around and begins to rise at large radius. The remaining galaxies have profiles that are nearly flat at all radii or fall, sometimes steeply, at all radii. (Figure 8, Appendix C)
- The outer $\sigma$ profile shapes correlate with galaxy luminosity. The most luminous 12 galaxies in our sample all have rising or nearly flat $\sigma$ profiles, whereas the less luminous ones show a wide variety of shapes.
- The outer $\sigma$ profiles also correlate with galaxy environment. Galaxies in groups and clusters tend to have rising or nearly flat $\sigma$ profiles, whereas nearly all (5/7) of our isolated galaxies have falling $\sigma$ profiles. Galaxies with larger halo masses have, on average, more steeply rising $\sigma$ profiles than galaxies with smaller halo masses, although the correlation is weaker with large scatter. (Figure 8, Figure 13)
- The wide variety of $\sigma$ profiles reported in this paper is a challenge for the standard power-law aperture correction schemes. Our sample roughly obeys, on *average* and when correcting from a central aperture of a few fibers to an aperture of $R_e$, the $\sigma \propto R^{-0.04}$ power law used for aperture corrections in Hyperleda (Jorgensen et al. 1995). However, the scatter is large and some galaxies with rising profiles will be corrected in the wrong direction. (Figure 6)

Finally, we are interested in the implications of our large sample of slow rotating ellipticals for dynamical mass modeling. The degeneracy between mass and velocity anisotropy prevents a straightforward equivalence between $\sigma$ and mass, but detailed measurements of the $h_4$ parameter of the LOSVD are one important ingredient for breaking that degeneracy.

- The luminosity-weighted average $h_4$, $\langle h_4 \rangle_e$, is positive for all 41 galaxies in our sample. The lower-mass galaxies show a range of values (from 0 to 0.05) while the higher-mass galaxies are limited to the upper end of that range. (Figure 11)
- We find a positive correlation between the radial gradient in $h_4$ and the outer radial gradient in $\sigma$ (quantified by $\gamma_2$), but there is no correlation between $\langle h_4 \rangle_e$ and $\gamma_2$. The correlation between the $h_4$ and $\sigma$ gradients may arise from gradients in circular velocity rather than from velocity anisotropy. (Figure 14)

We plan to report further investigations of stellar kinematics of MASSIVE survey galaxies in upcoming papers. We will use smaller unfolded bins to investigate the more detailed kinematic features such as twists, misalignments, and decoupled cores of a larger sample of MASSIVE galaxies. We will examine how galaxy kinematics relate to environment and assess if and how the kinematic morphology-density relation reported for lower-mass ETGs (Cappellari et al. 2011b; Fogarty et al. 2014) holds for massive ETGs. The diverse environments inhabited by the MASSIVE galaxies already indicate that slow rotators with $M^* \gtrsim 10^{11.5}\ M_\odot$ may be more common in low-density environments than the lower-mass slow rotators studied in the ATLAS$^{3D}$ survey. We will report more robust measurements of $R_e$ and $M_K$ of MASSIVE galaxies once our ongoing CFHT deep K-band imaging survey is complete.

## ACKNOWLEDGEMENTS

The MASSIVE survey is supported in part by NSF AST-1411945, NSF AST-1411642, HST-GO-14210, and HST-AR-14573. This research has made use of the Hyperleda database and the NASA/IPAC Extragalactic Database (NED) which is operated by the Jet Propulsion Laboratory, California Institute of Technology, under contract with the National Aeronautics and Space Administration. We thank Jeremy Murphy and Steven Boada for assistance with early observations, and Kate Reed, Stephen Chen, and Ben Kolligs for assistance with early data reduction.

## APPENDIX A: COMPARING CENTRAL VELOCITY DISPERSION TO LITERATURE VALUES

Figure A1 compares our values for the central $\sigma$ of each galaxy with literature values. We use $\sigma$ from the central fiber of the galaxy as our central $\sigma$, and the fiber radius of 2 arcseconds corresponds to an aperture of about 0.6 to 1 kpc depending on the distance of the galaxy. This is only approximate, as the center of the galaxy may not fall exactly in the center of a fiber. Values taken from Hyperleda (Paturel et al. 2003) are averages of available literature values which have been aperture corrected to 0.595 kpc according to the prescriptions of Jorgensen et al. (1995). The SDSS fiber size (Bolton et al. 2012) is also very similar, so we do not do any additional aperture corrections for values taken from the NSA. Note that the Hyperleda aperture corrections assume a radial $\sigma$ profile proportional to $R^{-0.04}$, which we have seen in Section 5.2 is not true for many of our galaxies. Nevertheless, we see very good agreement for most galaxies.

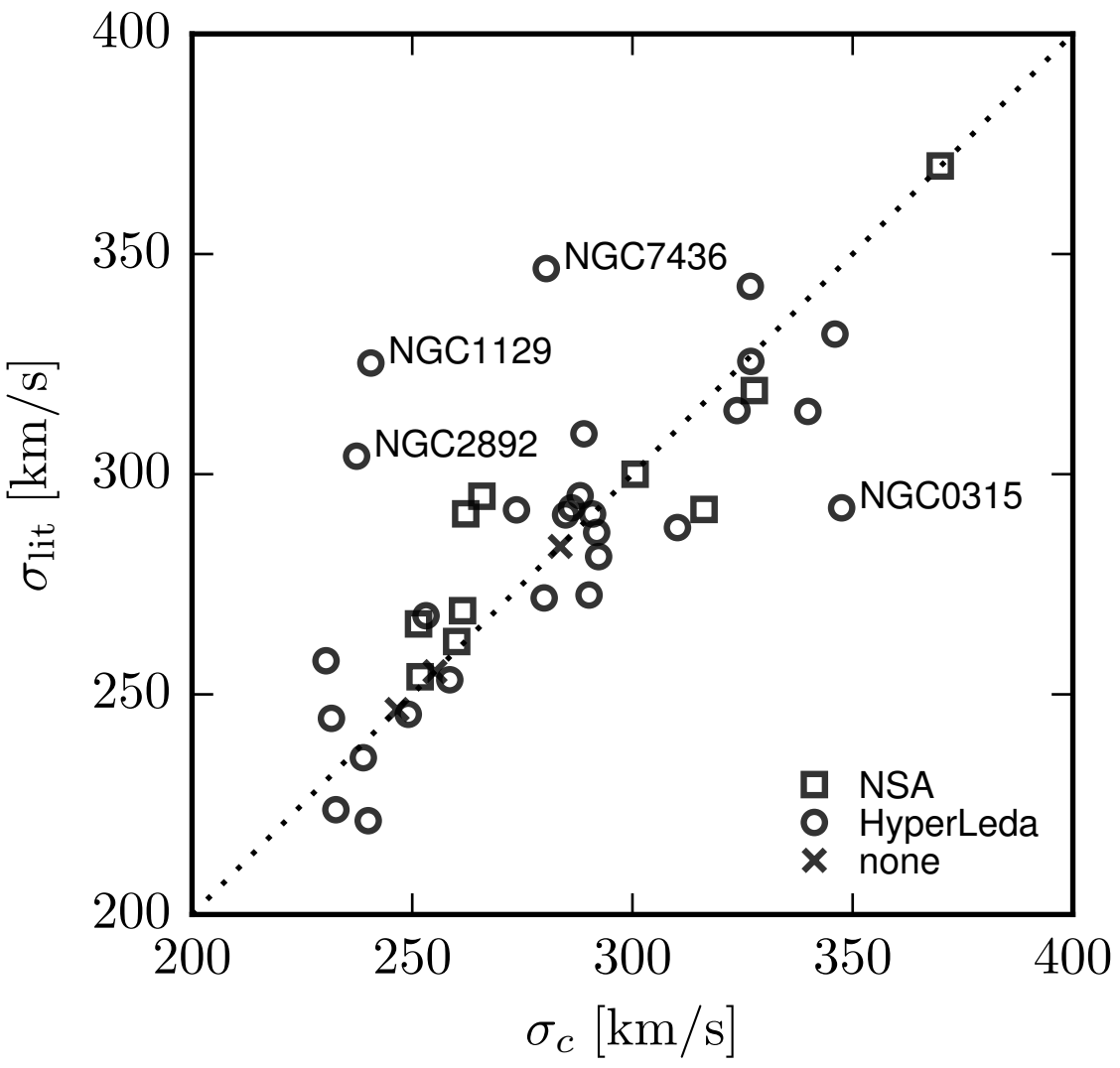

**Figure A1.** Comparison of our central $\sigma$ with literature values. Literature values are taken from Hyperleda (Paturel et al. 2003). Where available, the NSA values are shown instead (Bolton et al. 2012). The dotted line shows the one-to-one line. Individual galaxies with the worst agreement are labeled.

Only five galaxies show a difference greater than 30 km/s (a fractional difference of $\sim$ 10 to 15%) between our $\sigma_c$ and the literature values. Those galaxies are labeled in Figure A1: NGC 0315, 1129, 2256, 2892, and 7436. Paper II of the MASSIVE survey (Greene et al. 2015) compared central fiber $\sigma$ values for a larger subsample of MASSIVE galaxies to the HET catalog (van den Bosch et al. 2015) and found good agreement, so we do not repeat that comparison for this sample. We do note that four of the six above outliers are contained in the HET catalog, which finds $\sigma$ values much closer to ours than the Hyperleda values: NGC 0315 (325 km/s), 1129 (230 km/s), 2892 (273 km/s), and 7436 (313 km/s). One galaxy (NGC 0315) shows a very large range of values in the Hyperleda catalog, from 260 km/s to 360 km/s. Two of the galaxies (NGC 2256, 2892) have only one value listed in the Hyperleda catalog, both coming from the same dataset.

## APPENDIX B: COMPARING INDIVIDUAL GALAXIES WITH EXISTING KINEMATICS

Six galaxies are in common between the MASSIVE and ATLAS$^{3D}$ surveys (Ma et al. 2014). Of these, only NGC 4472 (M49) in the Virgo cluster is in the high-mass subsample studied in this paper. A comparison of the kinematics for all bins in the ATLAS$^{3D}$ and MASSIVE surveys is shown in the left panel of Figure B1. We include two additional common galaxies – NGC 5322 (middle panel) and NGC 5557 (right panel) – from our lower-mass sample for comparison. The agreement between the two surveys is excellent for all three galaxies. The MASSIVE results generally show less scatter at a given $R$ and cover two to five times farther in radius.

## APPENDIX C: DETAILS OF $\sigma$ PROFILE FITS

In this appendix we present the details of our fits to the radial $\sigma$ profiles. We parameterize the fit as follows:

$$\sigma(R) = \sigma_0 2^{\gamma_1-\gamma_2} \left(\frac{R}{R_b}\right)^{\gamma_1} \left(1 + \frac{R}{R_b}\right)^{\gamma_2-\gamma_1} \quad \text{(C1)}$$

where $\gamma_1$ gives the power law slope at small radius, $\gamma_2$ gives the power-law slope at large radius, and $R_b$ gives the break radius. This is similar to the Nuker fit for galaxy surface brightness profiles (Lauer et al. 1995), and is normalized such that $\sigma(R_b) = \sigma_0$. We emphasize that this particular fitting function is simply a convenient choice for quantifying the overall rise and/or fall of $\sigma$ with radius, and is not motivated by any physical reasoning.

Figure C1 illustrates the effect of each parameter in the fitting function, and the substantial degeneracies among parameters for typical values in our range of data. The most persistent degeneracy is between $R_b$ and $\gamma_2$, which cannot be broken effectively for most individual galaxies. For any given value of $\gamma_2$, it can be effectively made flatter or steeper by varying $R_b$ appropriately. We find that fixing the value of $R_b$ to 5 kpc for all galaxies does not have any impact on the quality of the fit. Note that on the other hand, fixing $\gamma_2$ and leaving $R_b$ free does impact the quality of fit slightly, and has the more concrete problem of resulting in wildly varying best-fit values for $R_b$. The reasonable best-fit values of $\gamma_2$ (which stay between $\pm 0.4$) in the case of fixed $R_b$ are much more convenient for quantifying the shape of the profiles.

Fixing $R_b$ to 5 kpc leaves us with two shape parameters, $\gamma_1$ and $\gamma_2$. These are still somewhat degenerate, even in profiles with a clear break between falling at small $R$ and rising at large $R$, and for profiles that fall monotonically at all radius the degeneracy becomes much worse. We account for this degeneracy by classifying some galaxies as well-fit by a single power-law, fixing $\gamma_1 = \gamma_2$ and thus also rendering $R_b$ moot. We classify galaxies as "well-fit" by the single power-law if the improvement in $\chi^2$ per degree of freedom between single and broken power-law is less than 0.3. Figure C2 illustrates this classification for three example galaxies, showing both the single and broken power-law fits. The top panel shows a galaxy well-fit by the single power law, and the middle panel shows a galaxy that clearly requires a break. The bottom panel shows a galaxy that nominally requires a break according to our $\chi^2$ classification, but the break is "reversed" to have $\gamma_2 < \gamma_1$ instead of $\gamma_2 > \gamma_1$. About 5 galaxies fall into this category, where the broken power law shows a *steeper* fall in $\sigma$ at large radius, but they are all relatively borderline cases where the single power-law still has a fairly good fit. For simplicity, we classify these as well-fit by the single power law, to give us only two major categories:

- Galaxies well-fit by a single power law, where $\sigma$ decreases (or is flat) at all radii.
- Galaxies requiring a broken power law, where $\sigma$ falls at small radius but begins to rise (or at least becomes flat) at large radii.

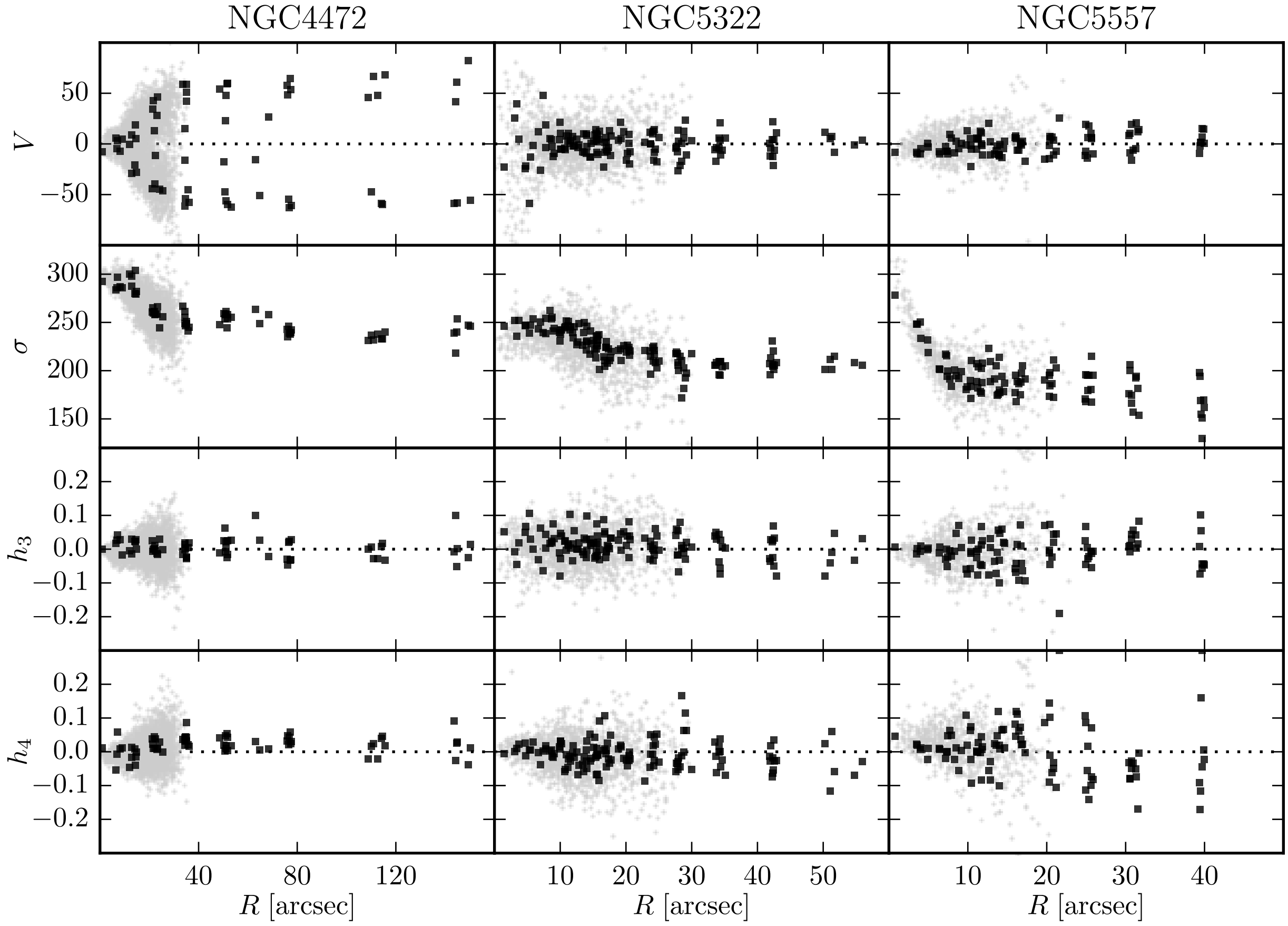

**Figure B1.** A comparison of ATLAS$^{3D}$ and MASSIVE kinematics for NGC 4472, 5322, and 5557. The radial profiles of four Gauss-Hermite velocity moments are shown (from top down). Each black square represents one bin from MASSIVE; each light gray point represents one bin from ATLAS$^{3D}$ .

Figure C3 shows all 41 galaxies in the plane of $\gamma_1$, $\gamma_2$. The two categories are evident, with 22 galaxies "well-fit" by a single power-law falling on the $\gamma_2 = \gamma_1$ line, and 19 galaxies requiring a break in the upper left quadrant. Note that the "reversed" galaxies (in blue) would nominally be located in the bottom right quadrant, because the best-fit broken power law for those cases has $\gamma_2 < \gamma_1$, but because we have chosen the single power-law as the best fit for these cases they have best fit parameters along $\gamma_2 = \gamma_1$. This also results in a clean separation between the two populations in $\gamma_2$, with all galaxies requiring a break having $\gamma_2 \gtrsim 0.07$ and all single power-law galaxies having $\gamma_2 \lesssim 0.07$. Because the split occurs at about $\gamma_2 \sim 0.07$, there is not a perfect separation between "falling" and "rising" profiles at large radius; for the most part, single power-law galaxies are falling at all radii while broken power-law galaxies rise at large radii, but there are some single power-law galaxies that rise (gently) at all radii. Because of the slight degeneracy in $\gamma_2$ and $\gamma_1$, shown in Figure C1, a positive $\gamma_2$ also does not guarantee strictly rising profiles at the largest radii we observe; if $\gamma_1$ is strongly negative, a positive $\gamma_2$ may merely result in a flattening of the profile. This degeneracy is also evident in Figure C3, where we see an anti-correlation of $\gamma_2$ and $\gamma_1$ due to the same effect: a more negative $\gamma_1$ requires a larger $\gamma_2$ to give the same results at large $R$. We do not attempt to disentangle this degeneracy further, and simply note that $\gamma_2$ is not entirely independent of the profile behavior at small $R$.

All galaxies have a $\chi^2$ per DOF around 4 or less, except for NGC 7426 ($\chi^2 \sim 7$, shown in red in Figure C3). Nonetheless, NGC 7426 appears by eye to be reasonably well fit by the single power law. Other galaxies with better $\chi^2$ may appear visually less well fit, due to more complicated radial profiles, but we do not attempt to treat them in any additional detail.

Two more galaxy properties are indicated in Figure C3 that are worth discussing here: whether our observations reach out to $R_e$ for the galaxy, and fast or slow rotator status. One might expect that for galaxies where our observations to not go out as far in radius (with sufficient signal to noise), we may be less likely to identify the point where $\sigma$ profiles begin to rise. The 6 galaxies with most limited observations, where the average radius of the outermost annulus is less than $R_e$, are circled in Figure C3. (Note that the total extent of the binned in these cases still goes out to $R_e$ or farther.) We find galaxies with limited observational extent nearly equally distributed among the break and non-break galaxies, suggesting that there is not an overwhelming bias in those galaxies with the most restricted observations. It is also worth noting that even for our galaxies with observations that go out farthest in radius, we are not generally sampling the profile very far beyond the apparent $R_b$. Our 7 fast rotators are similarly distributed throughout the $\gamma_1$-$\gamma_2$ parameter space, so there is no obvious relationship between the angular momentum content of a galaxy and it's dispersion profile.

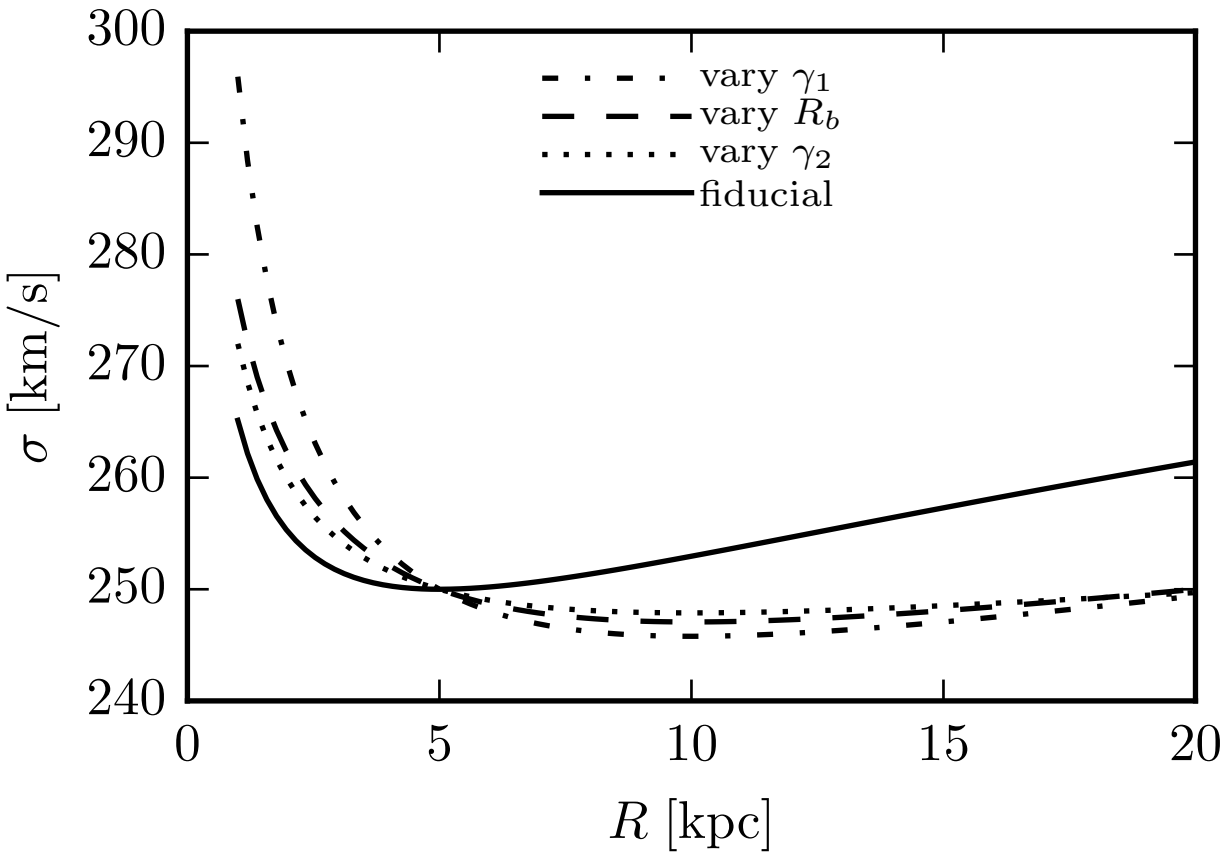

**Figure C1.** Shape of our fitting function for the $\sigma$ profiles, and the effects of varying different parameters. The fiducial curve uses reasonable parameter values for a galaxy showing a clear rise at large $R$. Each other curve varies only one parameter (and, if necessary, the normalization, to pin all curves to the same value at $R = 5$ kpc). The varied parameters are chosen to cause similar flattening of the profile at large $R$; note that varying any one of $\gamma_1$, $\gamma_2$, or $R_b$ can achieve almost identical effects. Numerical values of parameters are (for the fiducial curve): $\gamma_1 = -0.1$, $R_b = 5$, $\gamma_2 = 0.1$. Varied parameters are $\gamma_1 = -0.2$, $R_b = 10$, $\gamma_2 = 0.05$.

In principle, for this type of broken power-law fit, we would be able to break the degeneracy between $R_b$ and $\gamma_2$ by going farther out in radius; however, it is also extremely likely that we would find our choice of fitting function is not appropriate. It is precisely the limited radial extent of our data that allows us to choose a fairly arbitrary (i.e. not physically motivated) fitting function, while still achieving fairly good fits to the data.

## APPENDIX D: FULL GALAXY SAMPLE

Figure D1 through Figure D6 show kinematic results for all 41 MASSIVE galaxies of this paper. For each galaxy, the top row shows the 2D maps of $V$ and $\sigma$ (left two panels), as well as the fiber/bin map and galaxy image (right two panels); all four panels show the same field of view. The bottom row shows the radial profiles of $V$, $\sigma$, $h_3$, and $h_4$, including the fits to $\sigma$ and $h_4$ profiles used in Section 5.2 and Section 6.2.

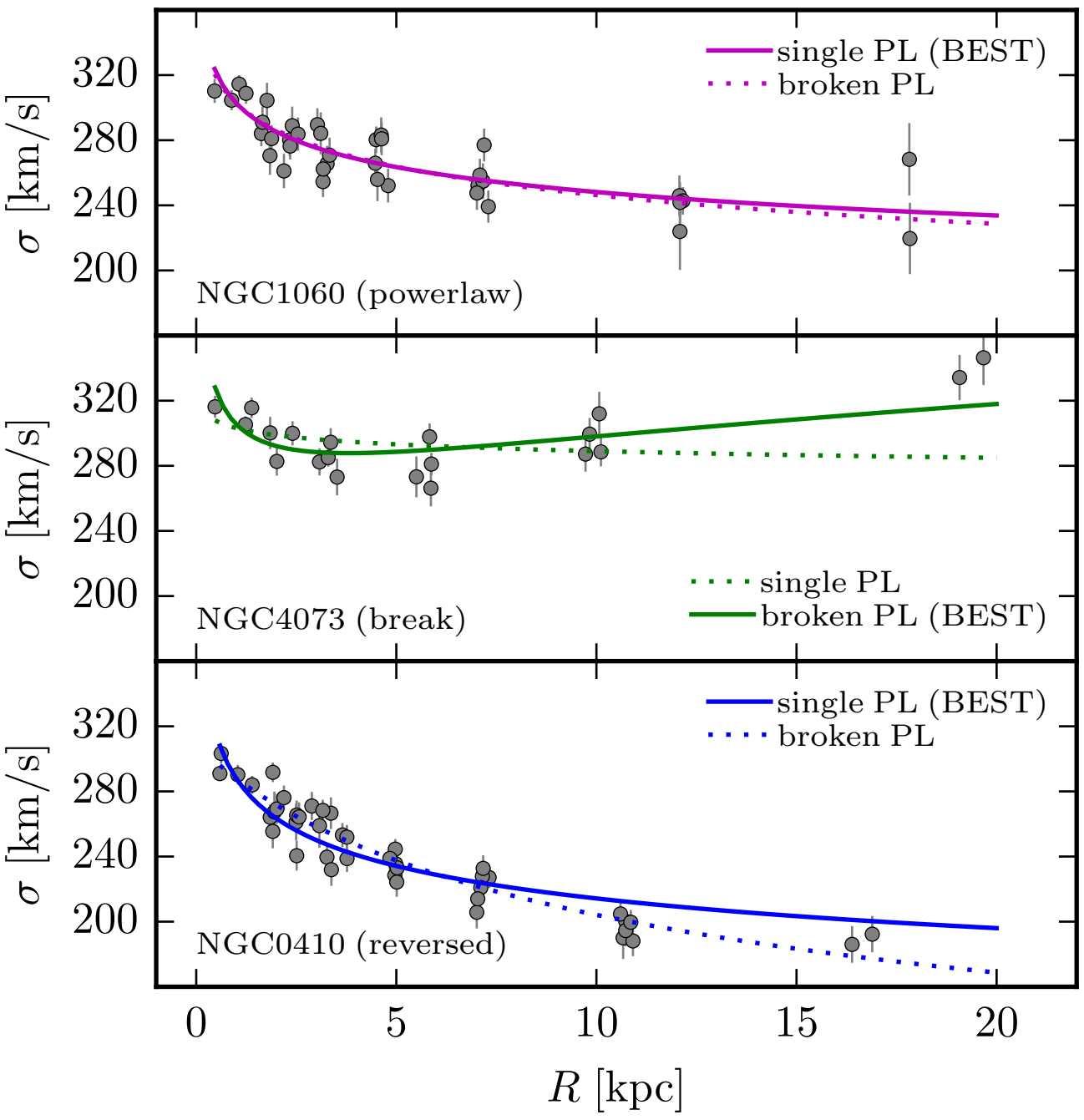

**Figure C2.** Example fits for $\sigma$ profiles of three galaxies, using both single and broken power-law. For each galaxy, the solid line shows the best fit curve, while the dotted line shows the alternate curve. The top panel shows a galaxy well-fit by a single power law. The middle panel shows a galaxy that requires a break. The bottom panel shows a galaxy that nominally requires a break according to the improvement in $\chi^2$ per degree of freedom, but the break is "reversed" with $\gamma_2 < \gamma_1$. About 5 galaxies fall into this category, and all are borderline cases, so we choose the single power-law as best fit for simplicity. Colors match those in Figure C3

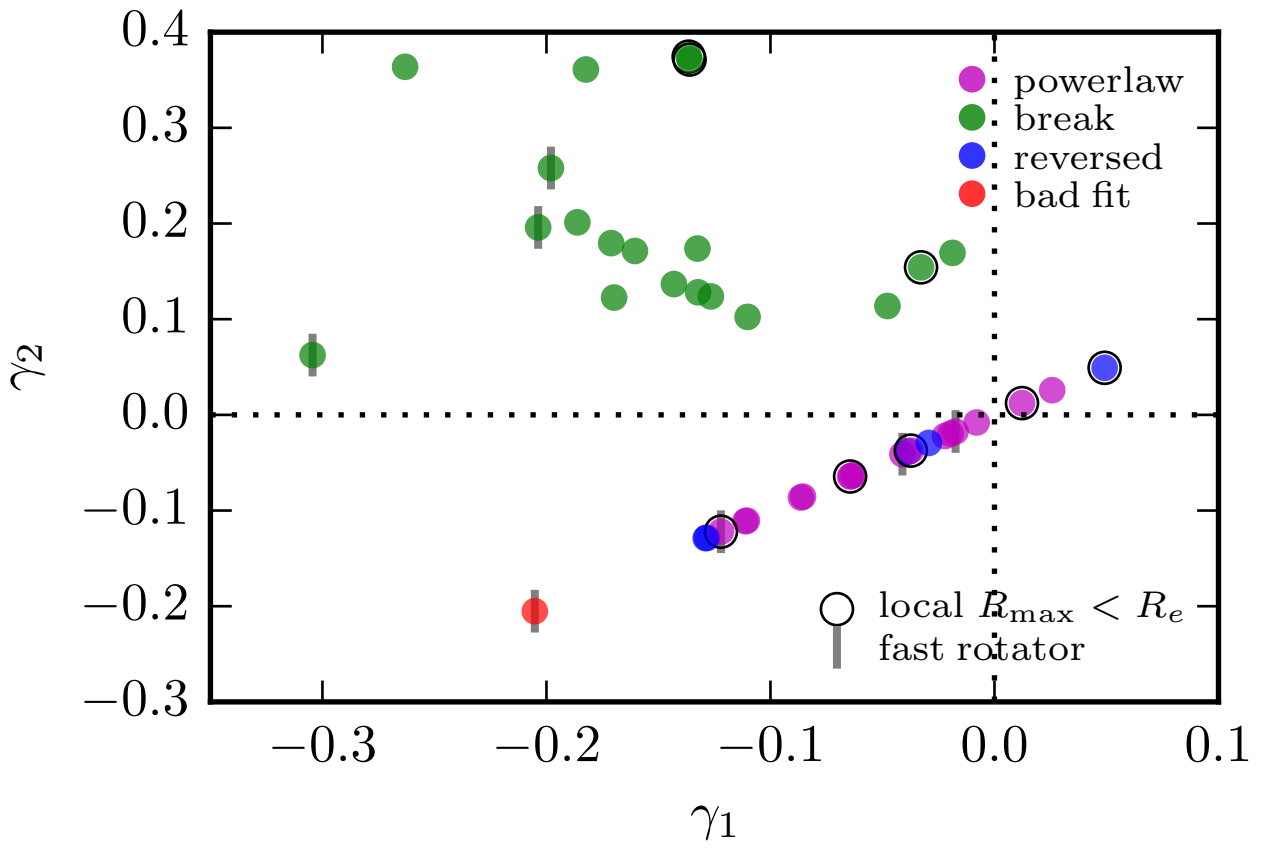

**Figure C3.** Best-fit parameters for the $\sigma$ profile of each galaxy. Colors correspond to the colors in Figure C2: magenta points are for galaxies well-fit by a single power-law, green points are galaxies requiring a broken power law due to rising $\sigma$ at large radius, and blue galaxies nominally require a broken power law that is "reversed" ($\gamma_2 < \gamma_1$) but are classified as well-fit by a single power-law for simplicity. The red point is NGC 7426, which has the worst $\chi^2$ per DOF of all the galaxies ($\chi^2 \sim 7$, nearly a factor of two worse than others), but nonetheless appears by eye to be reasonably well described by the single power law.

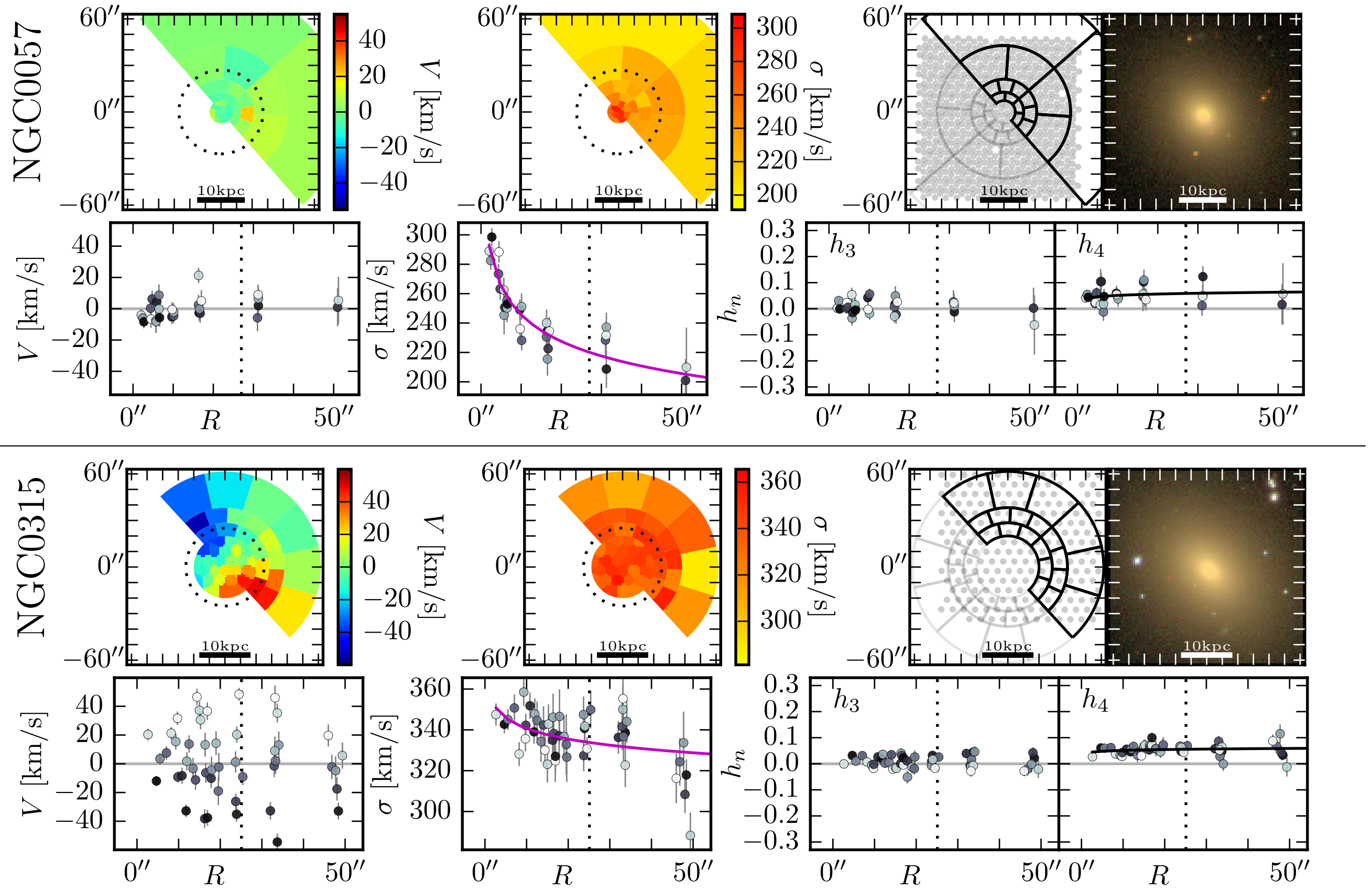

**Figure D1.** Kinematics results for NGC 0057 and 0315. The top row for each galaxy shows 2D maps ($V$, $\sigma$, fibers/bins, and optical image), all oriented such that North is up and West is to the right. The bottom row shows radial profiles ($V$, $\sigma$, $h_3$, and $h_4$). Ticks are always placed at intervals of 10″, and the dotted line (in both maps and radial profiles) indicates effective radius $R_e$. Images are taken from wikisky.org, using SDSS images where available and DSS2 images otherwise. The $\sigma$ vs radius panel also shows the best-fit $\sigma$ profile, color-coded as in Figure 8 (single or broken power law; see Appendix C). The $h_4$ vs radius panel also shows the best-fit $h_4$ profile (linear in $h_4$ vs $\log R$ space; see Section 6.2). The point color in the radial profiles corresponds to the angular location of the bin: black and white points correspond to 0 and 180 degrees from the PA respectively, and gray points correspond to bins near the minor axis.

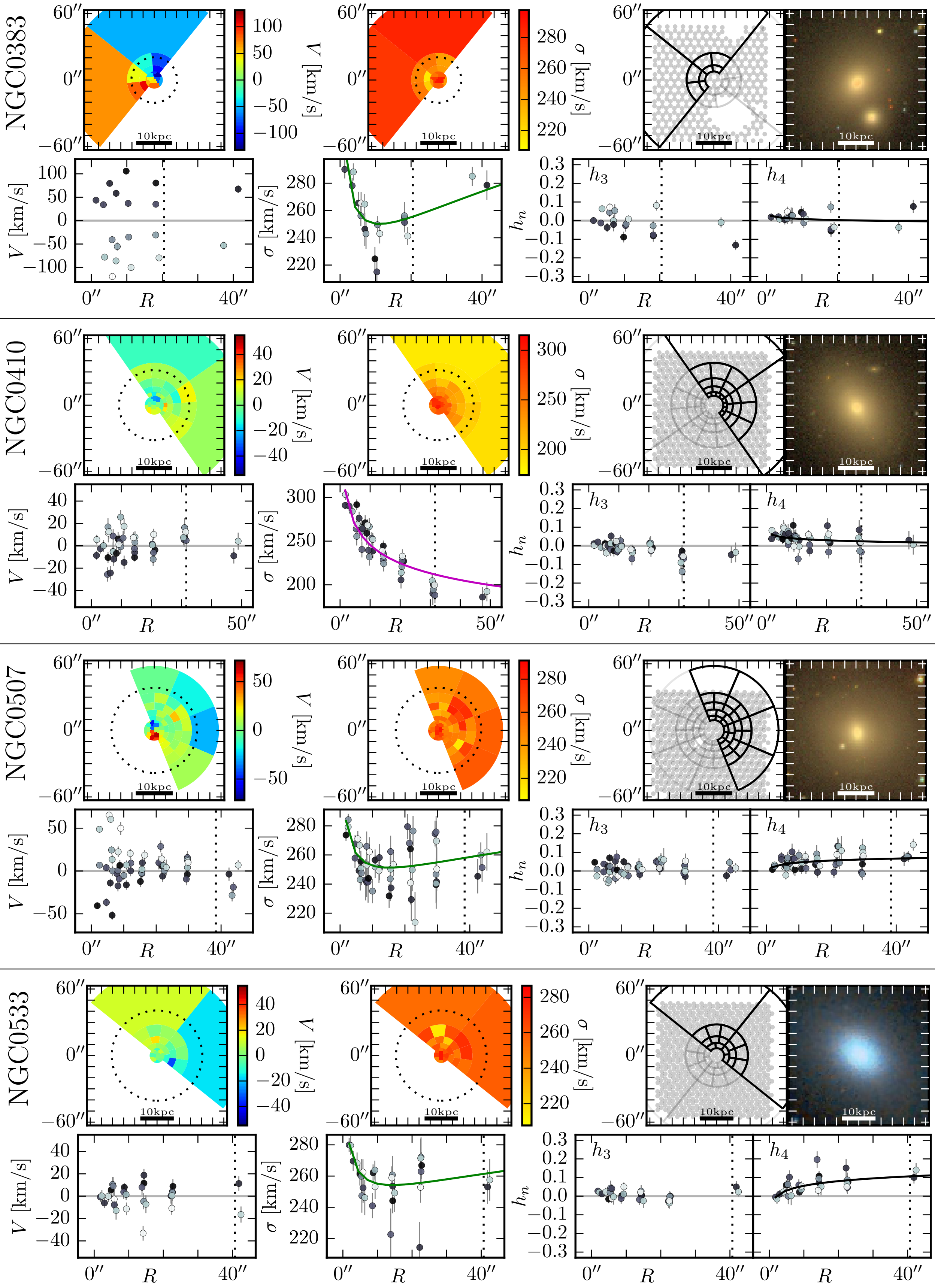

**Figure D2.** Kinematics results for NGC 0383, 0410, 0507, and 0533 (see Figure D1 for detailed caption).

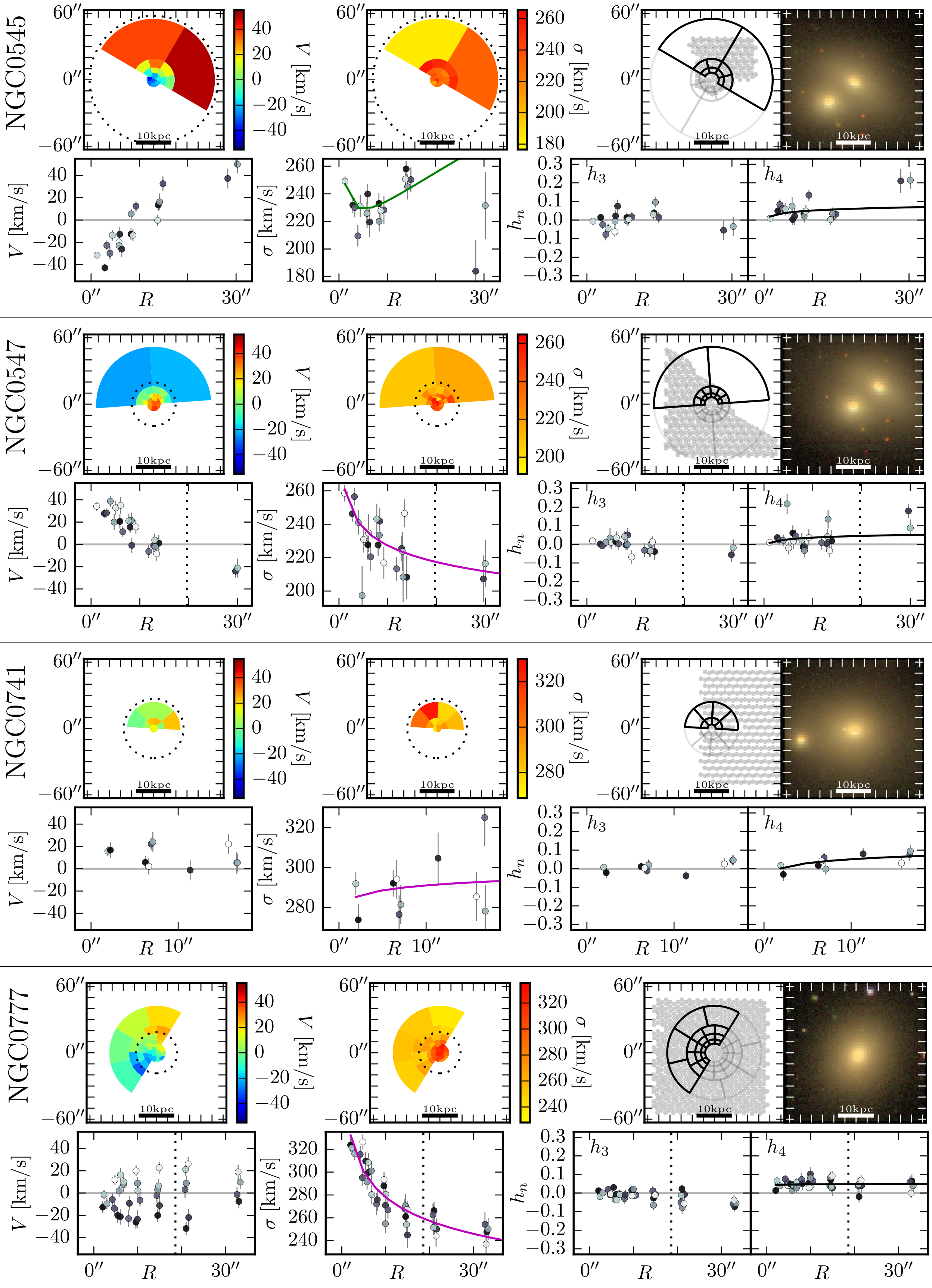

**Figure D3.** Kinematics results for NGC 0545, 0547, 0741, and 0777 (see Figure D1 for detailed caption).

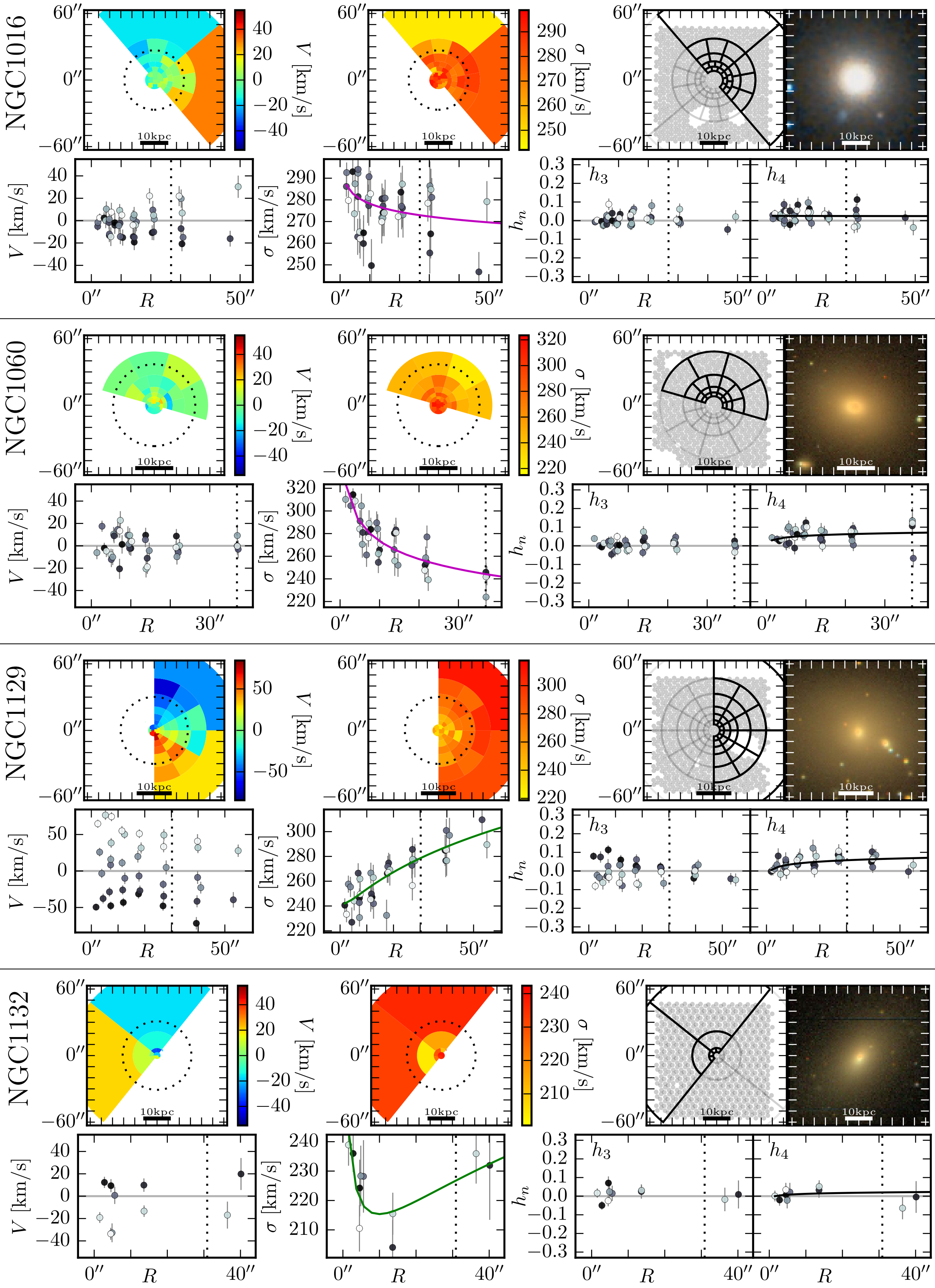

**Figure D4.** Kinematics results for NGC 1016, 1060, 1129, and 1132 (see Figure D1 for detailed caption).

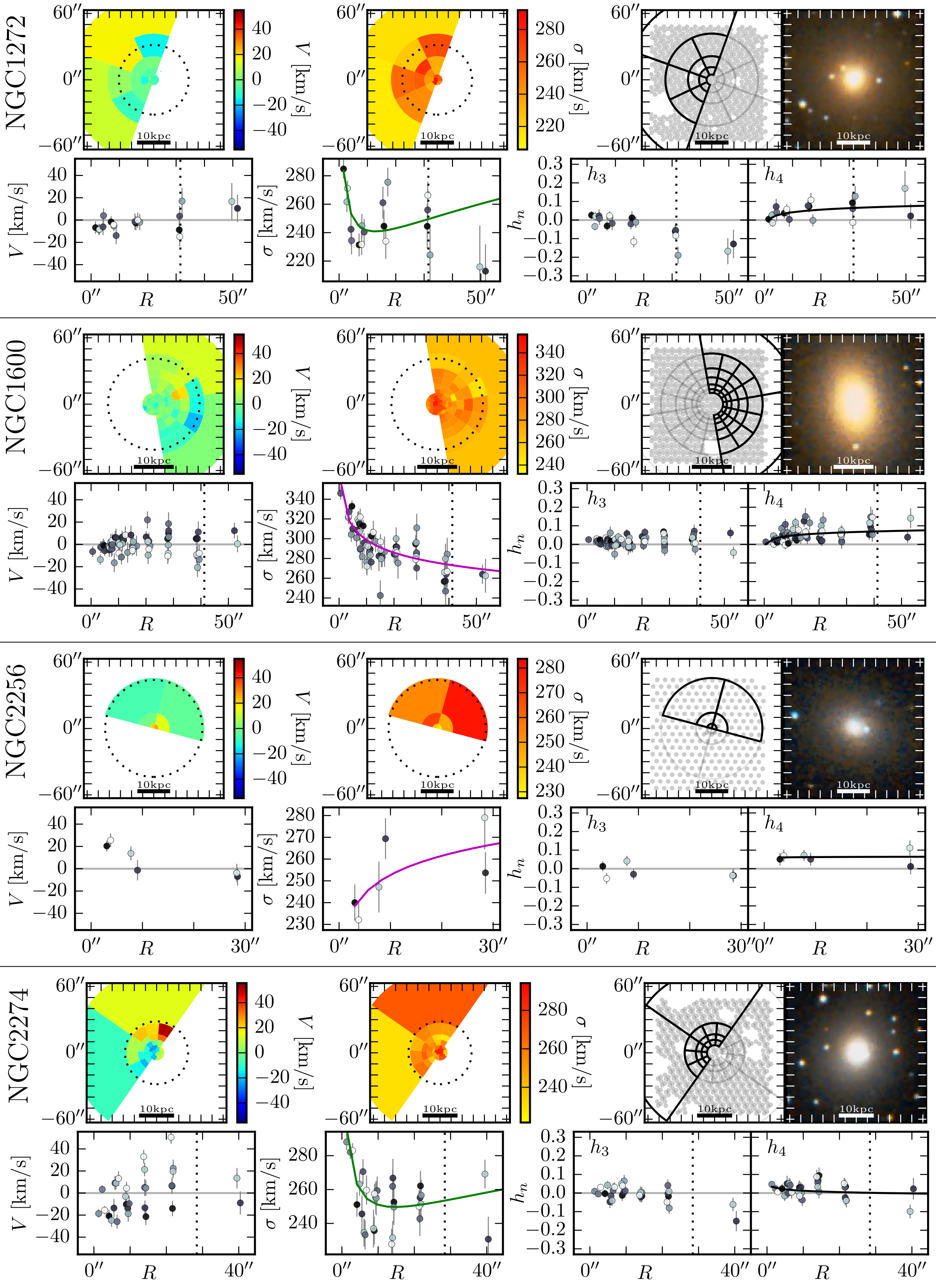

**Figure D5.** Kinematics results for NGC 1272, 1600, 2256, and 2274 (see Figure D1 for detailed caption).

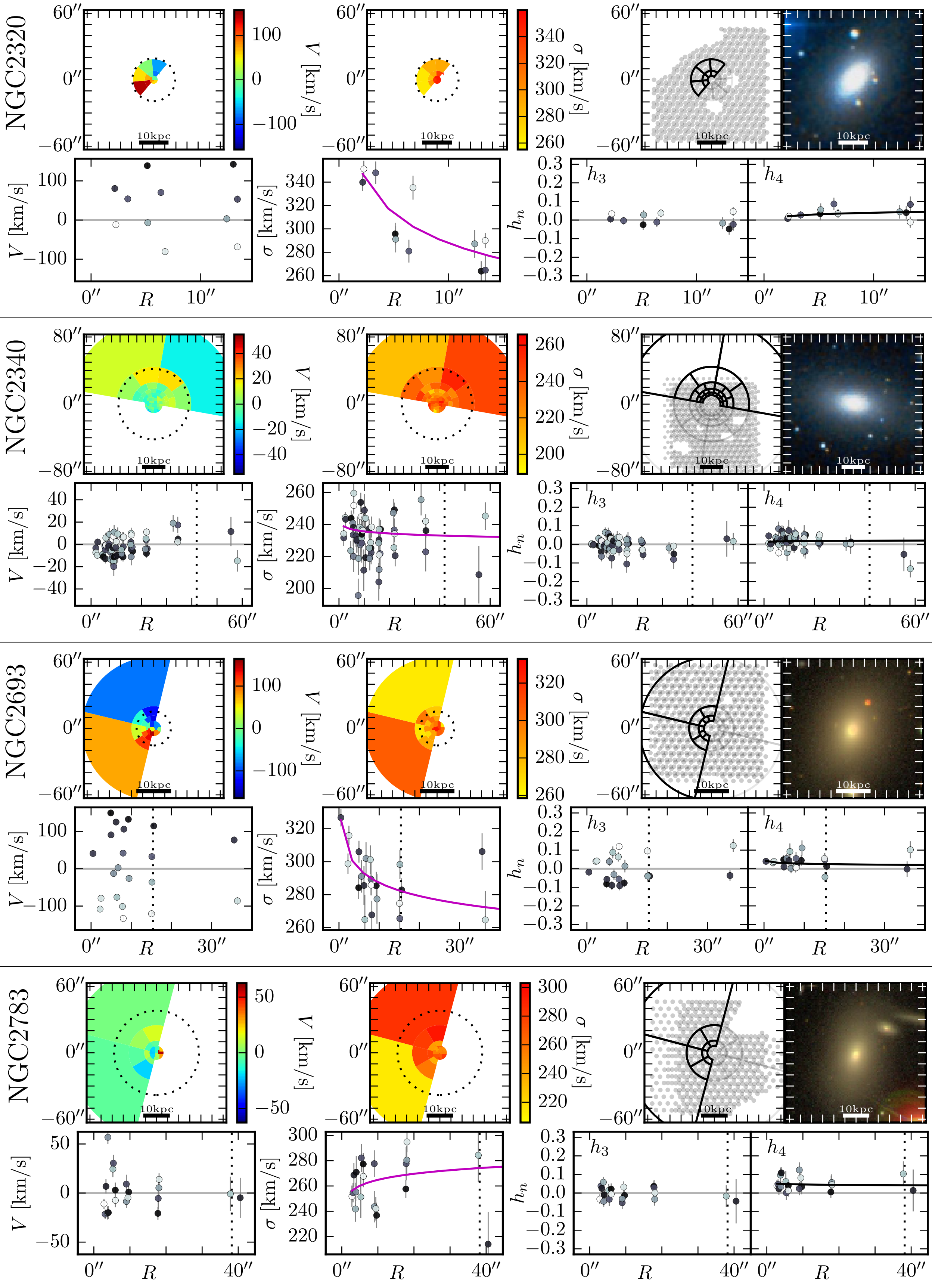

**Figure D6.** Kinematics results for NGC 2320, 2340, 2693, and 2783 (see Figure D1 for detailed caption).

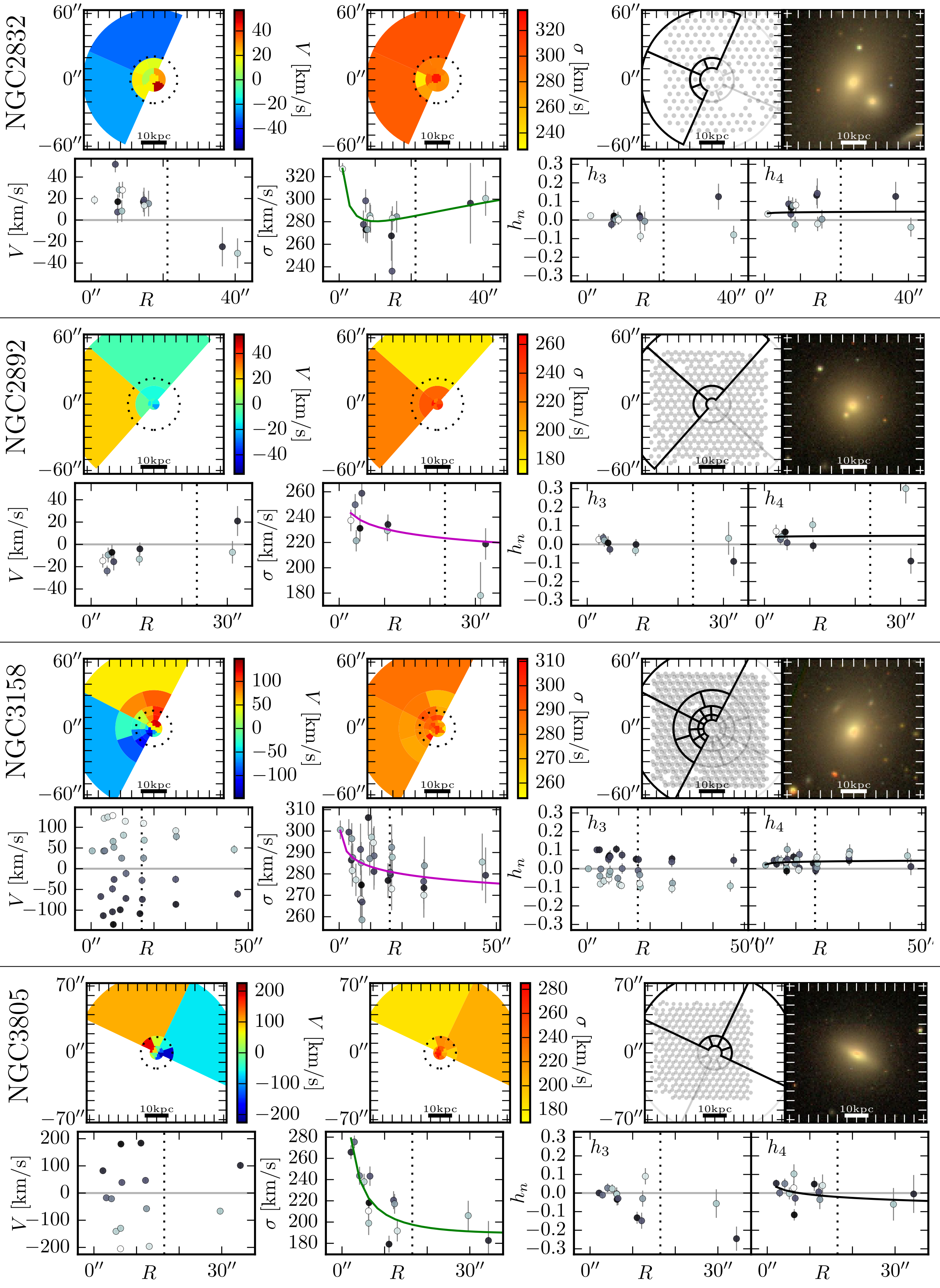

**Figure D7.** Kinematics results for NGC 2832, 2892, 3158, and 3805 (see Figure D1 for detailed caption).

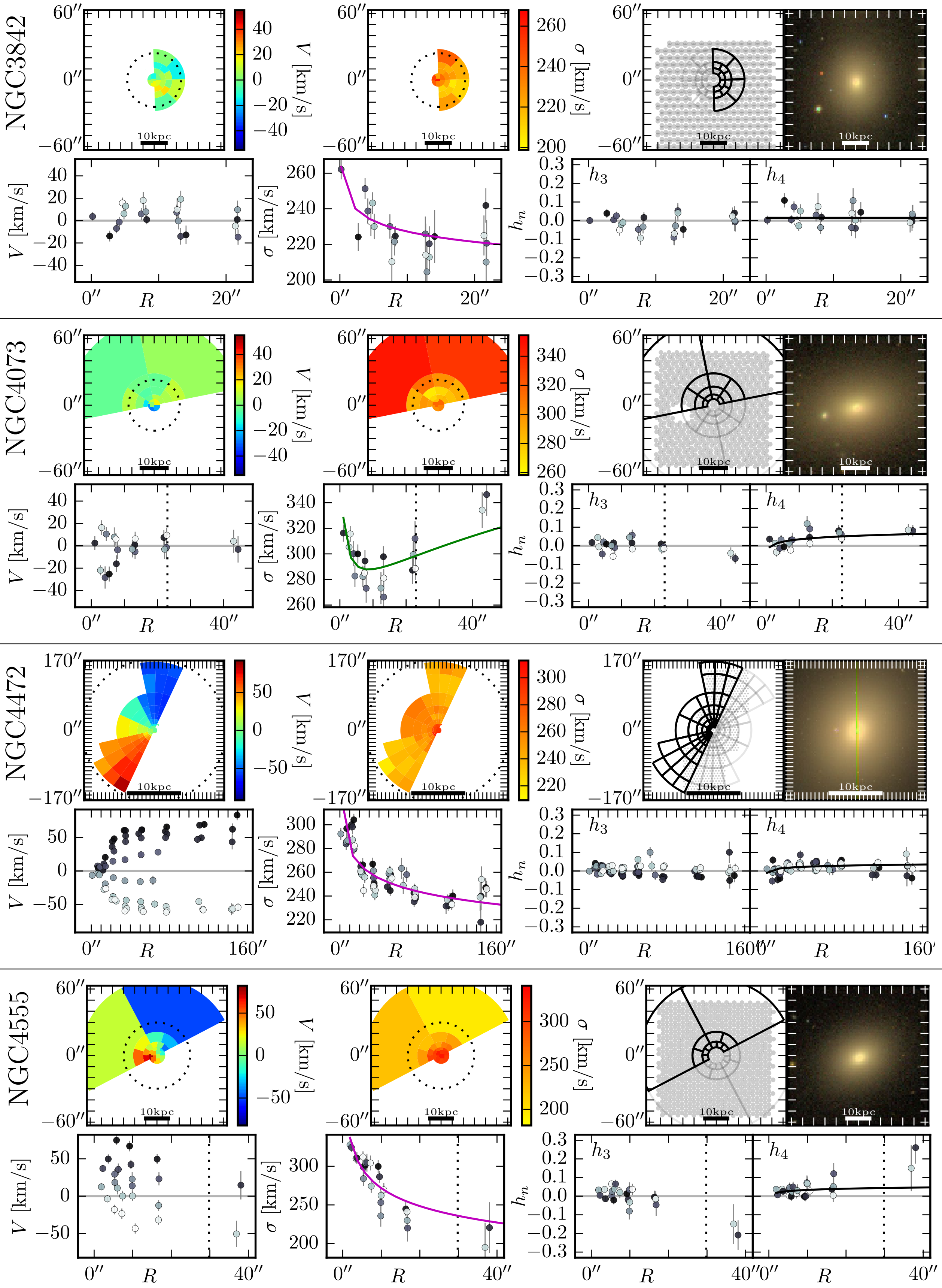

**Figure D8.** Kinematics results for NGC 3842, 4073, 4472, and 4555 (see Figure D1 for detailed caption).

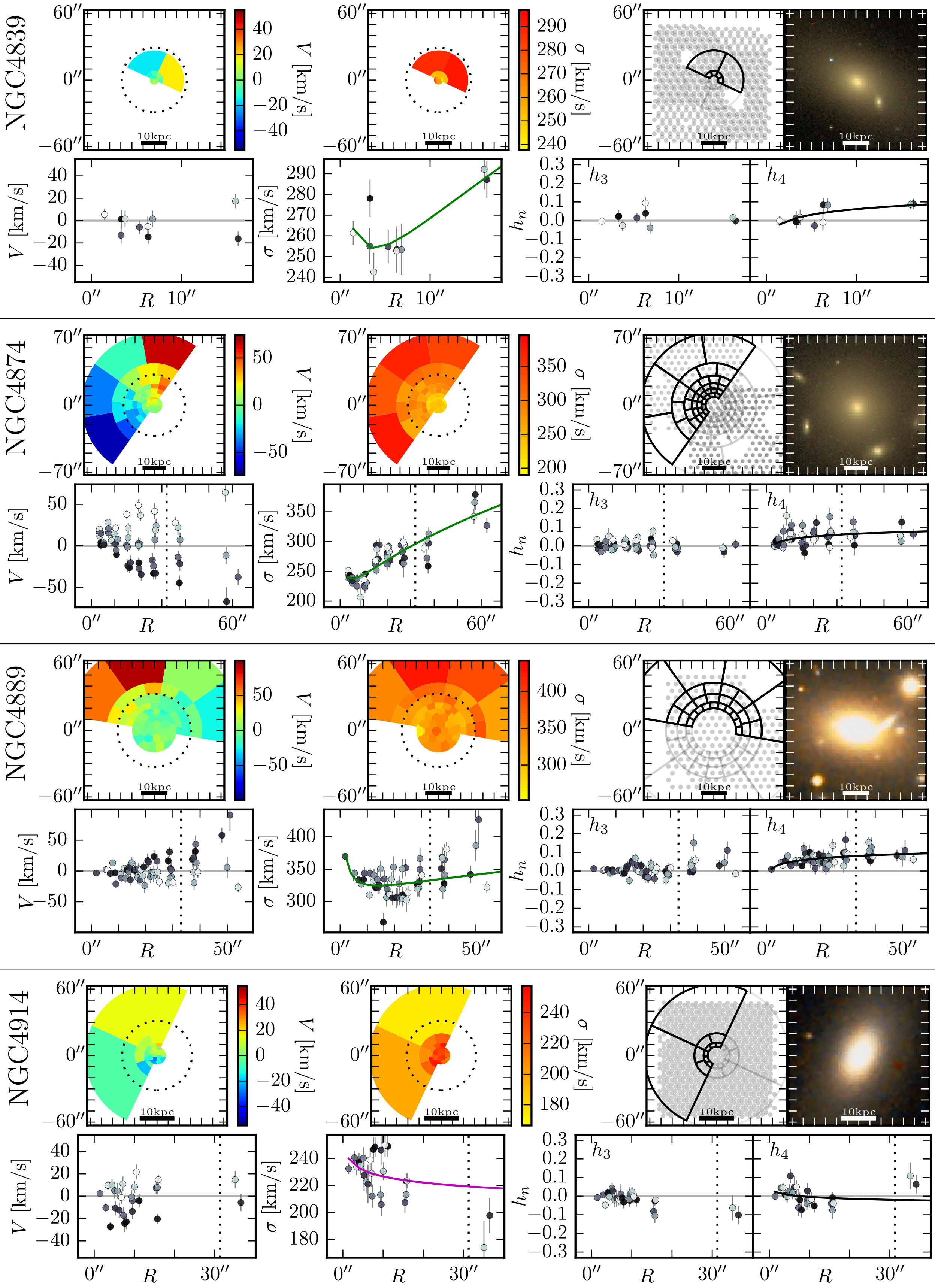

**Figure D9.** Kinematics results for NGC 4839, 4874, 4889, and 4914 (see Figure D1 for detailed caption).

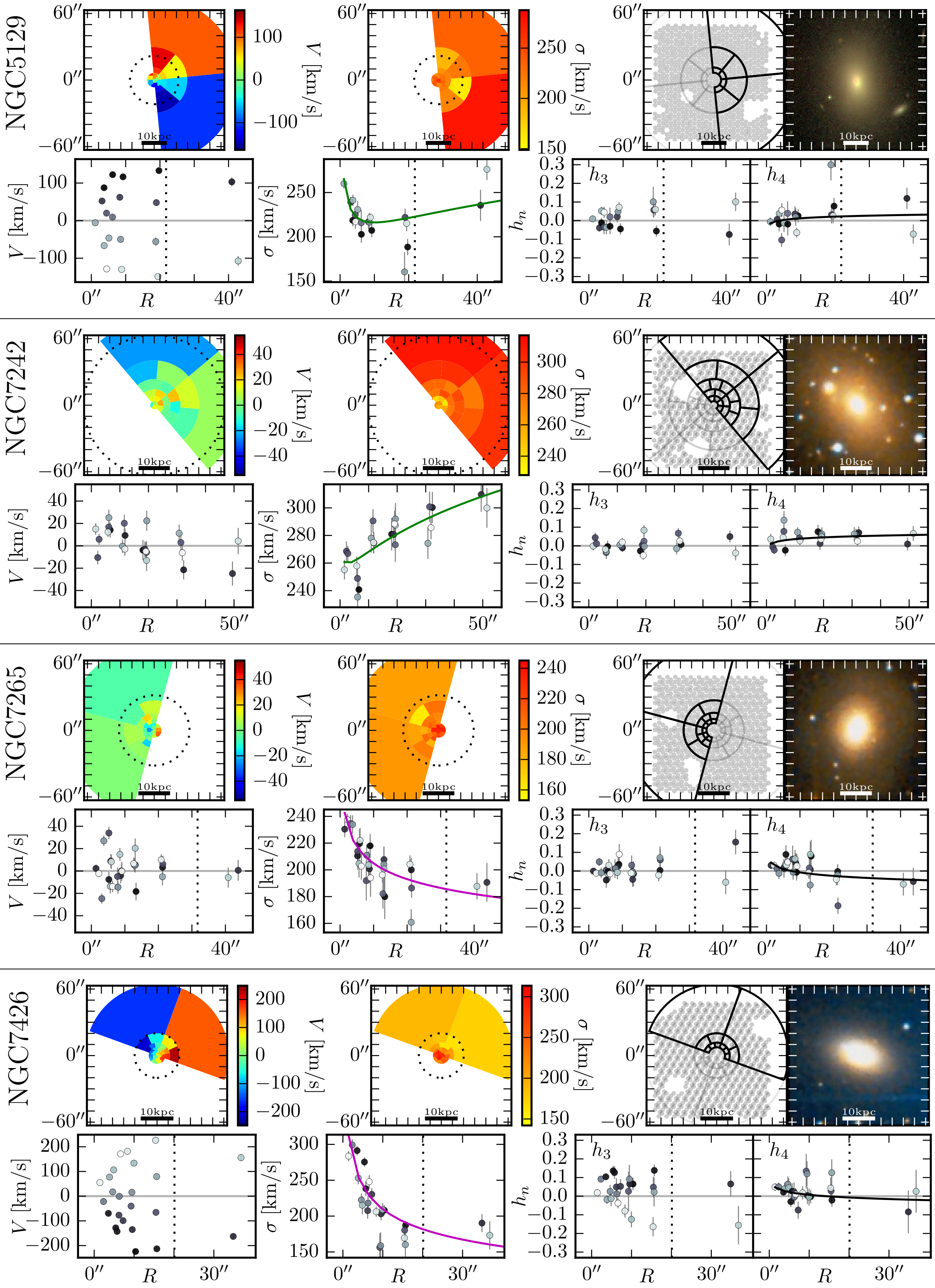

**Figure D10.** Kinematics results for NGC 5129, 7242, 7265, and 7426 (see Figure D1 for detailed caption).

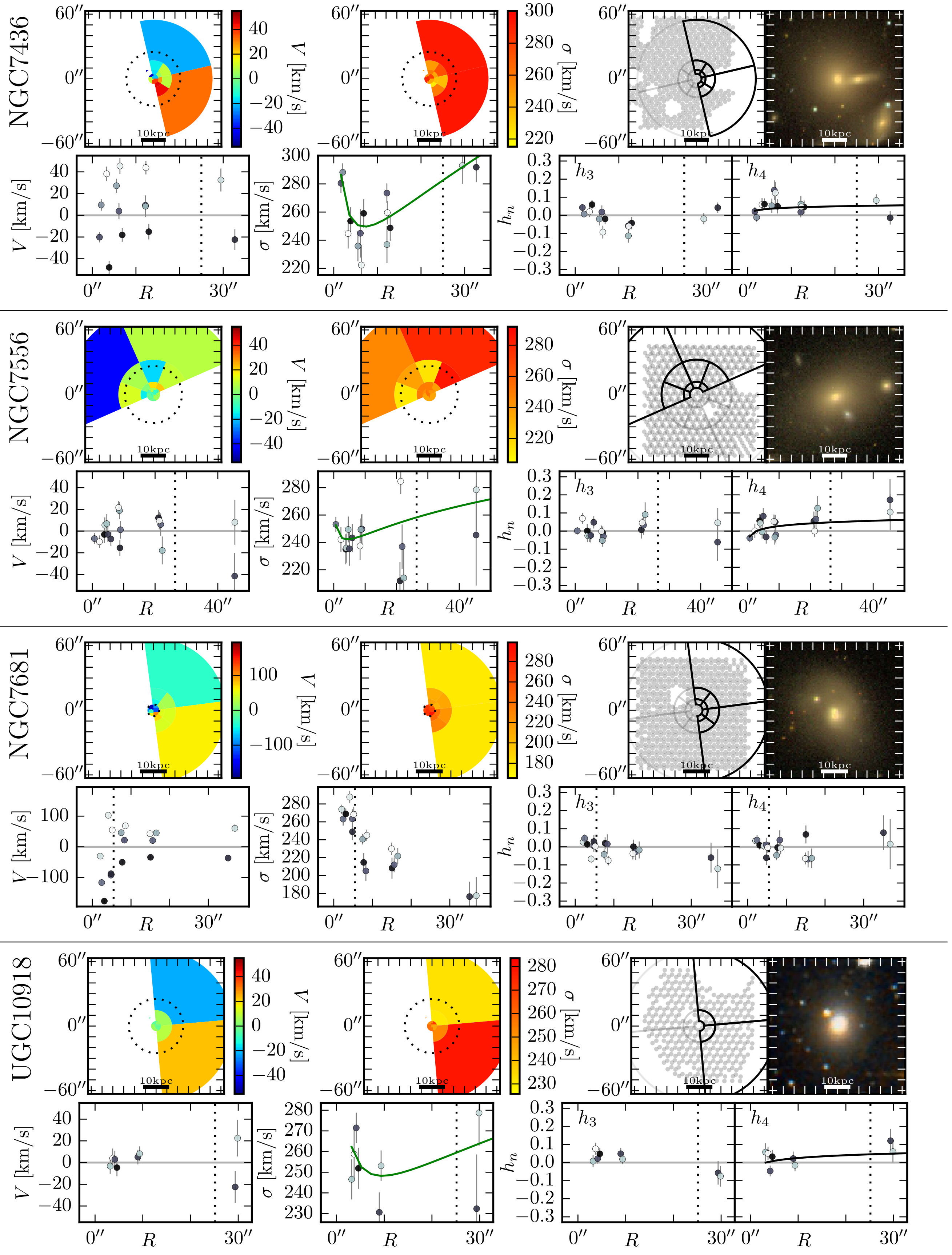

**Figure D11.** Kinematics results for NGC 7436, 7556, 7681, and UGC 10918 (see Figure D1 for detailed caption).